\documentstyle[aps,amstex,prd]{revtex}
\begin{document}
\makeatletter
\title{Non-Gaussian Fluctuations and Primordial Black Holes
 from Inflation}
\author{James S. Bullock\thanks{Electronic address: bullock@physics.ucsc.edu} 
 and Joel R. Primack\thanks{Electronic address: joel@physics.ucsc.edu}}
\address{Board of Studies in Physics, University of California, Santa
 Cruz, CA 95064}
\date{\today}
\makeatother
\maketitle

\begin{abstract}
We explore the role of non-Gaussian fluctuations in primordial black hole 
(PBH) formation and show that the standard Gaussian assumption, used in all 
PBH formation papers to date, is not justified.  Since large spikes in power 
are usually associated with flat regions of the inflaton potential, quantum 
fluctuations become more important in the field dynamics, leading to 
mode-mode coupling and non-Gaussian statistics.  Moreover, PBH production 
requires several sigma (rare) fluctuations in order to prevent premature 
matter dominance of the universe, so we are necessarily concerned with 
distribution tails, where any intrinsic skewness will be especially important.
We quantify this argument by using the stochastic slow-roll equation and a 
relatively simple analytic method to obtain the final distribution of 
fluctuations.  We work out several examples with toy models that produce
PBH's, and test the results with numerical simulations.  Our examples show
that the naive Gaussian assumption can result in errors of many orders of
magnitude.  For models with spikes in power, our calculations give sharp 
cut-offs in the probability of large positive fluctuations, meaning that
Gaussian distributions would vastly over-produce PBH's.  The standard
results that link inflation-produced power spectra and PBH number densities 
must then be reconsidered, since they rely quite heavily on the Gaussian 
assumption.  We point out that since the probability distributions depend
strongly on the nature of the potential, it is impossible to obtain results
for general models.  However, calculating the distribution of fluctuations
for any specific model seems to be relatively straightforward, at least in
the single inflaton case.        
\end{abstract}
\pacs{}

\newcommand{\gsim}{\mbox{\raisebox{-1.0ex}{$\stackrel{\textstyle >}
{\textstyle \sim}$ }}}
\newcommand{\lsim}{\mbox{\raisebox{-1.0ex}{$\stackrel{\textstyle <}
{\textstyle \sim}$ }}}
\newcommand{\beq}{\begin{equation}}
\newcommand{\eeq}{\end{equation}}
\newcommand{\beqa}{\begin{eqnarray}}
\newcommand{\eeqa}{\end{eqnarray}}
\newcommand{\lmk}{\left(}
\newcommand{\rmk}{\right)}
\newcommand{\dr}{\frac{\delta \rho}{\rho}}
\newcommand{\phidot}{\dot{\phi}}
\newcommand{\Zdot}{\dot{Z}}
\newcommand{\xdot}{\dot{x}}
\newcommand{\dZdx}{\frac{\partial Z}{\partial x}}
\newcommand{\dZdtau}{\frac{\partial Z}{\partial \tau}}
\newcommand{\dtaudt}{\frac{\partial \tau}{\partial t}}
\newcommand{\phiddot}{\ddot{\phi}}
\newcommand{\conh}{\frac{8 \pi}{3m_{pl}^{2}}}
\newcommand{\p}{\phi}
\newcommand{\td}{\tilde}
\newcommand{\Vtw}{\tilde{V}}
\newcommand{\Htw}{\tilde{H}}
\newcommand{\xtw}{\tilde{x}}
\newcommand{\xtwdot}{\dot{\xtw}}
\newcommand{\delH}{\delta_{H}}

\section{Introduction} 
\label{sec:intro}
Primordial black holes (PBH's) represent a link between quantum field theory 
and general relativity, both in their birth in inflation and their decay via 
Hawking radiation~\cite{hawk}.  Beyond the exciting physics of their 
formation and evolution, PBH's have the status of a dark matter (DM) 
candidate.  There is, in fact, renewed interest in this possibility, as 
Jedamzik~\cite{jed} has recently investigated whether PBH formation during 
the QCD epoch could explain the existence of the $\sim 0.5M_{\odot}$ MACHO's 
indicated by recent observations~\cite{alcock}.  Through the requirement that 
they not be over-produced, PBH's are also important as one of the few 
constraints on the inflationary Universe scenario~\cite{carrlid,guth2,linde2}.
Since these objects are an 
important potential link to the early universe, their formation should be 
examined in detail, especially if the standard Gaussian assumption about 
their formation may be producing errors of many orders in magnitude.

In order to determine the degree to which non-Gaussian fluctuations are 
important, we use a stochastic inflation calculation to investigate several 
PBH-producing toy models.  We solve for the probability distributions both 
analytically and numerically, and show that the Gaussian assumption is 
significantly flawed.  Specifically, for models associated with spikes in 
small-scale power, the Gaussian assumption over-estimates PBH
production by many orders of magnitude.  These results suggest that if some
model of inflation is in danger of being ruled out due to overproduction of 
PBH's (\textit{e.g.}, as considered by~\cite{guth2,linde2}), then looking 
closely for non-Gaussian behavior may serve to save the model, or at least 
alter the ranges of viable parameters.

We start with a heuristic discussion. Our intuition about non-Gaussian 
statistics for single inflaton models is fostered by the simple relation 
between the inflation-produced power spectrum and the field dynamics driving 
inflation.  Using this relation, we are able to make general statements
about the behavior of inflaton fluctuations in models which form PBH's, and 
see qualitatively why non-Gaussian statistics might be important.  In order 
to make this point, we briefly review the standard inflation scenario, set 
our notation, and use these results to illustrate why non-Gaussianity seems 
likely in PBH formation.

The inflationary Universe scenario~\cite{guth} provides the seeds of 
structure formation by linking initial density perturbations to quantum 
fluctuations in one or more scalar fields.   See~\cite{turner} for a review.
During inflation, the energy density of the Universe, $\rho$, becomes 
dominated by a scalar field potential, $V(\p)$, and the scale factor $R$  
expands superluminally ($R \sim t^{n}, n > 1$).  So length scales
of fluctuations grow more quickly than the horizon, eventually
passing out of causal contact, and only cross back inside the horizon
after the inflation epoch has ended.  In the chaotic scenario~\cite{linde}, 
the spatially homogeneous field, $\phi$, is initially displaced
 from the minimum of its potential and rolls downward with its
motion governed by the Klein-Gordon and Einstein equations  
\begin{eqnarray}
	\phiddot + 3H(\phi)\phidot & = & -V'(\phi), \\
	              H^{2} & = & \conh \rho = \conh (V(\phi) + 
	                        \phidot^{2}/2), 
\end{eqnarray}
where $m_{pl}$ is the Planck mass and $H$ 
is the Hubble parameter.  In most instances, $H\approx \sqrt{\conh}
{V(\phi)}^{1/2}$ and the ``slow-roll'' conditions 
\begin{eqnarray}
	  \left|V''\right| &\lsim& 24 \pi V/m_{pl}^{2}, \nonumber \\
	  \left|V'\right| &\lsim& \sqrt{48\pi}V/m_{pl} \nonumber 
\end{eqnarray}
apply.  The evolution of $\phi$ is then friction dominated:
\begin{equation}
	  \phidot \simeq -\frac{V'}{3H}, 
\end{equation}
and the universe continues to inflate until either of the slow
roll conditions breaks down at some value $\phi \equiv \phi_{end}$.       

As the physical size of fluctuations grows and they cross outside the
scale of the horizon during inflation, their amplitude is set by quantum
fluctuations in $\phi$.  Since the energy density 
$\rho$ is dominated by $V(\phi)$, we have  
$\delta \rho \sim V' \delta \phi \sim V'H$.  After inflation 
ends $R \sim t^{n}, n < 1$ and perturbations
begin to cross back inside the horizon.  The rms magnitude of 
$\delta\rho/\rho$ when a scale re-enters the horizon, $\delH$,
is simply related to 
its properties as it left the horizon. Using the gauge-invariant
variable $\xi \simeq \delta \rho/(\rho + p)$, we have~\footnote{The
expression for $\delta_{H}$ is correct up to 
factors of order unity which depend \textit{e.g.} on whether there is 
radiation or matter domination when the scale re-enters the horizon.}
\begin{eqnarray}
	\delH(L) =  \left(\dr\right)_{re-enter}
	       \simeq \left(\frac{\delta\rho}{\rho + p} \right)_{re-enter}
                =  \left(\frac{\delta\rho}{\rho + p} \right)_{out}
	       = \left.\frac{\delta\rho}{\phidot^{2}} \right|_{out}
	       \approx  \left.\frac{V^{3/2}}{V'm_{pl}^{3}} \right|_{out}
\end{eqnarray}
where $L$ is the co-moving scale of horizon crossing and the 
right hand side is evaluated when the fluctuation left
the horizon at $\phi = \phi_{out}(L)$. When a fluctuation
crosses outside the horizon it has physical size $H^{-1}(\phi_{out})$,
so to find the co-moving size today, we must scale by
the amount of expansion since that time: $L = [R_{o}/R(\p_{out})]
H^{-1}(\phi_{out})$.   We find the relation 
between $\p_{out}$ and the present-day length scale $L$ by 
first determining the amount of expansion that occurred during the rest
of inflation [$R(\p_{end})/R(\p_{out})$] and then using
entropy conservation to determine the amount
of expansion that took place up to now [$R_{o}/R(\p_{end})$].
We know $\dot{R}/{R} = H(\phi) \Rightarrow 
d\ln R = Hd\phi/\phidot$ during inflation, so the inflationary expansion was
quasi-exponential:
\begin{eqnarray}
L(\phi_{out}) &=& \left(\frac{R_{o}}{R_{end}}\right)
	\left(\frac{R_{end}}{R(\p_{out})}\right)H^{-1}(\phi_{out})
	= \frac{R_{o}}{R_{end}}\exp(N(\phi_{out}))H^{-1}(\phi_{out})
\end{eqnarray}
where the number of ``e-folds'' between exiting the horizon and
the end of inflation, $N(\p_{out})$, is given by
\begin{eqnarray}
   N(\phi_{out}) &=&  \int_{\p_{out}}^{\p_{end}} \frac{H(\p)d\p}{\phidot}  
  \approx  \frac{8\pi}{m_{pl}^{2}} \int_{\p_{out}}^{\p_{end}} 
	\frac{V(\p)d\p}{V'(\p)}. 
\end{eqnarray} 
For the expansion since the end of inflation, we will here assume
a ``quick re-heat'' approximation and use
$R_{o}/R_{end} \approx [3.17](T_{end}/T_{o}) \approx
 V(\phi_{end})^{1/4}/0.85K$, where the numerical term arises due to the
effective number of relativistic species.

The crucial point from the above discussion is that every region of the
 inflaton potential is mapped directly to some scale of the fluctuation
 spectrum using equations (4)--(6).  For typical model parameters, $L \sim
1$ Mpc corresponds to $N(\p) \approx 51$ 
and $L \sim 1$ pc corresponds to $N(\p) \approx 37$.  We invert
such relations to determine $\p({\rm pc}),\p({\rm Mpc})$, etc.

Specific characteristics 
of the density spectrum over a range of wavelengths then allow us to make 
general statements about the behavior of the potential in some region of 
$\p$ space.  As we shall see in Sec. II, PBH production requires a 
specially-tuned fluctuation spectrum: the fluctuation amplitude on small 
scales must increase over its value on the COBE scale by a factor of 
$\sim 10^{3}$ in order to produce an appreciable fraction of PBH's.
It is this that makes non-Gaussianity important in these models.

Hodges $\textit{et al.}$~\cite{hodges}
 have examined the question of non-Gaussianity
in some detail\footnote{For examinations of 
specific models using the stochastic inflation
approach see $e.g.$~\cite{yvm}.} 
and showed that non-Gaussianity is negligible in 
single-inflaton models with spectra meeting the large-scale structure
requirement $\delH \sim 3\times10^{-5}$.   
  However, models that produce PBH's
must have density $\delH \gsim 0.01$
over some range of small-scale wavelengths ~\cite{carrlid,carrhawk}.   
In order to get a qualitative
feel for why non-Gaussian fluctuations may be important in these models,
consider a background perturbation expanded in terms of the usual
creation and annihilation operators $a_{k}^{\dag},a_{k}$
\begin{eqnarray}
	\p(x,t)& =& \p_{cl}(t) + \delta\p(x,t) \nonumber \\
	\delta\p(x,t) & = & \frac{1}{(2\pi)^{3}}
	  \int d^{3}k\left[a_{k}\delta\p_{k}e^{ikx} + h.c.\right]. \nonumber
\end{eqnarray} 
Inserting the perturbed field into the
free-field Klein-Gordon equation we obtain
\begin{equation}
	\delta\phiddot_{k} - \frac{k^2}{R^{2}}\delta\p_{k}
	+ 3H\delta\phidot_{k} + V''\delta\p_{k} = 0,
\end{equation}
where we have expanded to first order in $\delta\p$ and matched small
terms.  Gaussian statistics are preserved to first order since
in the linear approximation all Fourier modes remain separate.

As $\delta\p$ becomes larger, however, the linear
approximation breaks down and terms of order $\delta\p^2$ 
become important.  Any non-linearity implies mode-mode coupling 
and introduces non-Gaussianity.  So the possibility of a non-Gaussian 
distribution of fluctuations increases with $\delta\p$.
Moreover, PBH production \textit{requires} fluctuations of several $\sigma$ 
above $\delta_{rms} \equiv \delH$ in order to keep their production 
relatively rare (see Sec. II for a discussion).  So we may expect 
non-Gaussian statistics to play a role in PBH production in inflation.

A much stronger effect becomes apparent if we examine the general nature of 
a potential $V(\p)$ that produces PBH's.  Since PBH production requires 
a large increase in power on small scales, expression
 (4) requires that $\phidot$
(or similarly $V'$) become extremely small over some region of 
the potential.  In these cases, writing down a decomposition like
equation (7) makes no sense: every term is ``small,'' so matching
terms of equivalent size does not simplify the equation.  In other
words, quantum fluctuations become important in the dynamics of
$\p$. We must keep $\delta\p$ linked to $\p$, and in the slow-roll
approximation we have
\begin{eqnarray}
	\frac{d}{dt}(\p + \delta\p) = -\frac{V'(\p + \delta\p)}
					{3H(\p + \delta\p)}. \nonumber
\end{eqnarray}
The evolution becomes extremely non-linear and non-Gaussian 
effects become important.

In what follows, we investigate the question of non-Gaussian statistics
and PBH production in detail.  In Sec. II we present the basics of 
PBH creation and the implications for the inflation-produced power spectrum.
We review stochastic inflation in Sec. III and present the
method of deriving probability distributions that we will use in our examples.
In Sec. IV, we present three examples of inflationary scenarios
which create PBH's in significant numbers, and solve for the distribution
of fluctuations both analytically and numerically.  We reserve Sec. V 
for conclusions and speculation.

\section{Primordial Black Holes and Constraints on Inflation}
\subsection{Formation}
A PBH forms when a collapsing over-dense region is large enough to 
overcome the opposing force of pressure and falls within its
Schwarzschild radius~\cite{carrhawk}.  We review the basics of this process 
for a radiation-dominated universe, where the sound speed is given by 
$c_{s} = c/\sqrt{3}$, and the pressure is $p = c_{s}^{2}\rho =
\rho/3$.  Consider a spherically symmetric\footnote{Note that
the spherical assumption is well justified ~\cite{doroshkevich} for
many--$\sigma$ ``rare'' fluctuations in a \textit{Gaussian} random
field.  This result may actually be incorrect for non-Gaussian fluctuations
and should be explored further.} 
region with 
density $\td{\rho}$ greater
 than the background density $\rho$.  The high-density 
region is governed by the positive curvature ($K > 0$) Friedmann equation
\begin{eqnarray}
	\td{H}^{2}(\td{t}) = \conh\td{\rho}(\td{t}) - 
		\frac{1}{\td{R}^{2}(\td{t})}
\end{eqnarray}
while the background space evolves as
\begin{equation}
	H^{2}(t) = \conh\rho_{b}(t). 
\end{equation}
Following the standard approach, 
choose initial coordinates $t = \td{t} = t_{i}$ such that the initial
expansion rates in each region are equal $\td{H} = H = H_{i}$.
Using (8) and (9) we see that the initial size of 
the perturbation $\delta_{i}$
in terms of $\td{R}(t_{i}) \equiv R_{i}$ is given by
\begin{equation}
	\delta_{i} \equiv \frac{\td{\rho}_{i} - \rho_{i}}
	 {\rho_{i}} = \frac{1}{H_{i}^{2}R_{i}^{2}}.
\end{equation}
Before the over-dense region breaks away from
the expansion, $\td{R} \sim \td{t}^{1/2}$~\cite{harrison}, 
and we can estimate the time, $t_{c}$,
when the perturbed region stops expanding ($\td{H}(t_{c}) = 0$):
\begin{equation}
	t_{c} \simeq \delta_{i}^{-1}t_{i}
\end{equation}
and the corresponding expansion factor at this time
\begin{equation}
	\td{R}(t_{c}) = R_{c} \simeq \delta_{i}^{-1/2}R_{i}.
\end{equation}
The perturbed region continues to contract only if it contains enough
matter to overcome the pressure, 
$R_{c}$ $ \gsim $ $R_{Jeans} = c_{s}t_{c} \sim \frac{1}{\sqrt{3}}t_{c}$.
With this condition met, a black hole will inevitably form:
we need not worry about vorticity or turbulence
interfering with the formation process since the region will fall inside
its Schwarzschild radius soon after turn-around ($R_{S} 
= 2GM \simeq 2G\rho_{c}R_{c}^{3} \simeq R_{c}^{3}t_{c}^{-2} 
\gsim \frac{1}{3}R_{c}$).  The requirement for
PBH formation is then
\begin{equation}
	\frac{1}{\sqrt{3}}t_{c} \lsim R_{c} \lsim t_{c}
\end{equation}
where the upper bound prevents the region of collapse from being
larger than the horizon scale, leading to the formation of
a separate universe~\cite{carrhawk}. 
Now, dividing (13) by $t_{c}$, we obtain a condition on
$R_{c}/t_{c} \simeq R_{i}/t_{i} \simeq \delta_{i}^{1/2}$.
This expression scales like a constant with time, so it is convenient
to evaluate it at horizon crossing, $R=t$.  We then have our final
condition for perturbations at horizon crossing
\begin{equation}
	\frac{1}{3} \lsim \delta \lsim 1.
\end{equation}
Thus, given an initial fluctuation with $\delta$ satisfying (14)
at horizon crossing, we expect to form a PBH with
mass $M$ corresponding roughly to the horizon mass at that time
(numerical results agree ~\cite{nadezhin}).

\subsection{Probing the Power Spectrum}
Although quite elegant in solving the problems of the standard
big bang, inflation provides us only with a framework of ideas
and not with any exact predictions for universal evolution.
Depending on the scalar potential, number of fields, etc.,
inflation can produce a wide variety of fluctuation spectra 
\textit{e.g.}~\cite{hodges,hodgesblum,lindebook,kolbvarenna95}
and even the
possibility of a low $\Omega$ Universe~\cite{lindeastroph}.
Because of the freedom in inflationary models, any means of 
constraining a particular scenario becomes extremely important.
The Cosmic Background Explorer (COBE) measurements 
and large-scale structure observations
indicate $\delH \sim 3\times10^{-5}$ on large scales 
($\sim10^{3}$Mpc).  These results probe only a small region of the 
inflation potential $V(\phi)$ and serve mainly to fix the value of the 
inflaton's dimensionless coupling constant ($\textit{e.g.}$ for $V(\p) 
= \lambda\p^{4}$, we need $\lambda \sim 10^{-14}$).
Over-production of PBH's limits the value of $\delH$ over many 
decades of ``small'' length scales ($1$Mpc -- $10^{-10}$pc) and avoidance
of such over-production,
coupled with the COBE data, provides a powerful constraint on
inflation~\cite{novikov,carrlid}.

Limits on black hole production are usually quantified by the parameter
$\beta$, defined to be the initial mass fraction of PBH's
\begin{equation}
	\beta \equiv \frac{\rho_{BH}^{*}}{\rho_{TOT}^{*}}
	\approx \int_{\frac{1}{3}}^{1}P(\delta)d\delta,
\end{equation}   
where (*) means we evaluate the ratio at the time of formation
and the limits of integration are directly from (14).  The probability
distribution of density fluctuations, $P(\delta)$, is usually assumed
to be Gaussian, with the power spectrum giving us $\sigma \equiv
\delta_{rms} = \delH$ at any particular scale.  For the rest of this section we will assume
$P(\delta)$ is of Gaussian form
\begin{equation}
	P(\delta) = \frac{1}{\sqrt{2\pi}\sigma}
		\exp(-\frac{\delta^2}{2\sigma^2}),
\end{equation}
and use the results as a gauge for any alternate distributions we find.

There are two criteria for limiting the initial abundance of
PBH's (see \textit{e.g.}~\cite{carrlid}).  First, gravitational effects: the mass density of PBH's must
not over-close the universe, $\Omega_{PBH} \equiv 
(\bar{\rho_{PBH}}/\bar{\rho_{tot}})|_{today} < 1$.   And second, Hawking
radiation from decaying PBH's must not disrupt any well-constrained
physics ($\textit{e.g.}$ primordial nucleosynthesis).
Since PBH's with $M \lsim 10^{15}$g
will have decayed before today, gravitational effects for this 
mass range are non-existent.  Similarly, larger mass PBH's have not 
had time to decay so Hawking radiation does not come into play.
Thus, the physics of constraining PBH's is broken up into two mass regions:
\begin{enumerate}
\item Gravitational constraints: $M \gsim 10^{15}$g
\item Hawking radiation constraints: $M \lsim 10^{15}$g
\end{enumerate}
We discuss each in turn briefly.

\bigskip
\noindent\textbf{Limits from Gravitation: $\Omega_{PBH} < 1$}

\noindent We are interested in constraining the parameter $\beta$, the 
$\textit{initial}$ density fraction of PBH's, using our limit on the 
$\textit{current}$ 
fraction: $\Omega_{PBH} \lsim 1$.
We simply need to scale the density ratio with time. We consider
PBH formation in the radiation dominated era, thus $\Omega_{PBH} \sim
R$ until matter-radiation equality, after which the density fraction
remains constant.
If $R^{*}$ is the epoch of formation we have
\begin{equation}
  \beta \equiv \Omega_{PBH}^{*} = \left(\frac{R^{*}}{R_{eq}}\right)
   \Omega_{PBH} \lsim 10^{-5}\left(\frac{t^{*}}{s}\right)^{1/2}
\end{equation}
where the inequality demands $\Omega_{PBH} \lsim 1$ and
the time $t^{*}$ corresponds to $R(t^{*}) = R^{*}$.    
Since the black holes formed are roughly on the scale of the horizon
at the time of collapse, we can write (15) in terms of mass
scale using the following expression for horizon mass
\begin{equation}
 M_{H} \approx 10^{5}\left(\frac{t}{s}\right)M_{\odot}
\end{equation}
which yields
\begin{equation}
	\beta \lsim 10^{-17}\left[\frac{M}{10^{15}g}\right]^{1/2}.
\end{equation}
The constraint becomes weaker for larger masses since higher mass
PBH's form later and the density, $\rho_{PBH}$, has a shorter
time to grow relative to the total density.  We now use (15),(16) to
map the $\beta$ values to limits on $\delH$.  
Note that since $\beta$ is quite small, 
we depend very much on the tail of our assumed 
Gaussian distribution.  For a $10^{15}$ g PBH 
we have $\beta \lsim 10^{-17}$ and the limit becomes $\delH
\lsim 0.04$.  Similarly, for $M \sim 1M_{\odot}$, $\delH 
\lsim 0.06$.

\bigskip
\noindent\textbf{Limits from Radiation}

\noindent Hawking's theory of spontaneous black hole 
evaporation~\cite{hawk},~\cite{hawk2}
predicts that black holes should emit particles at a characteristic
temperature $T_{H} \approx 10^{13}(M/{\rm g})^{-1}$ GeV, which becomes
more important with decreasing black hole mass.
The lifetime of decay is $\tau 
\approx 10^{-27}(M/{\rm g})^{3}$ s~\cite{page}, and is similarly 
dependent on the black hole mass,
with the evaporation speeding up as the mass dwindles.  PBH's of
different initial masses evaporate at different times and their particle
emission must not disrupt any well-understood physics 
associated with the time of their decay.  We summarize a few
constraints below.  For more complete reviews 
see~\cite{novikov,carrlid,carrnew}.

\begin{itemize}

\item \underline{$M \sim 10^{14}-10^{15}$ g}:  Black holes in this
mass range evaporate after recombination and must not produce 
a $\gamma$-ray density greater than the cosmic background~\cite{pageh}.
The constraint here is quite strong: PBH's could have at most $10^{-8}$ 
of the critical density today.  We then have $\beta \lsim 10^{-25}$
which corresponds to $\delH \lsim 0.03$.  We emphasize that
since PBH production needs $\delta \sim 1/3$, formation in this mass range
corresponds to a $10\sigma$ effect assuming a Gaussian distribution of 
fluctuations.

\item \underline{$M \sim 10^{11}-10^{13}$ g}: For these masses,
evaporation occurs before recombination but the emitted radiation
does not attain equilibrium due to the small value of the baryon
to photon ratio.  This effect will distort the background spectrum
unless $\beta \lsim 10^{-18}(\frac{M}{10^{11}{\rm g}})$~\cite{zeldovichstarb}.
For a typical mass, the result implies $\delH \lsim 0.04$,
and a Gaussian $8\sigma$ fluctuation
is required for PBH formation.
 
\item \underline{$M > 10^{10}$ g}: This mass range marks the possibility
of affecting big bang nucleosynthesis.  For example, emitted photons from 
the evaporation must not photodissociate $D$.  This limit
gives~\cite{lindley}: $\beta \lsim 10^{-21}(\frac{M}{10^{10}{\rm g}})^{1/2}$
and
corresponds to $\delH \lsim 0.03$ for $M \sim 10^{10}$ g. 
PBH production again corresponds to a $10\sigma$ Gaussian fluctuation.
\end{itemize}

\noindent From the above discussions we see that PBH over-production provides  
limits on the initial spectrum of perturbations on mass scales small
compared to the horizon mass $\sim 10^{55}$g.  We want to translate the
above limits as a function of mass to limits as a function of co-moving
length scale $L$.  For lengths that cross inside the horizon before
matter-radiation equality ($L \lsim 13h^{-1}$ Mpc) the time of
horizon crossing is
\begin{equation}
 t_{H} \simeq 3\times10^{8}(L/{\rm Mpc})^{2}{\rm s}
\end{equation}
Together with (18) we have the following relation
\begin{equation}
 M_{H} \sim 30\left(\frac{L}{{\rm pc}}\right)^{2}M_{\odot}
\end{equation}
with $1M_{\odot}$ corresponding to $L \sim 0.2$ pc and $10^{15}$ g
corresponding to $L \sim 2\times10^{-10}$ pc.

Since COBE fixes $\delH \sim 3\times10^{-5}$ on large scales, PBH constraints
serve to limit the power spectrum from having a spike in power at
some small length scale, or from having an excessively steep blue spectrum.
Since $\delH$ must be $\gsim 0.01$ in order for PBH formation to be important,
we are concerned with factors of $\sim 10^{3}$ enhancement of the
spectrum at small scales.  We point out again that such a power
spectrum is inherently associated with potentials that can
make our Gaussian assumption invalid.  This non-Gaussian tendency
coupled with the fact that we are concerned with many-$\sigma$ tails of 
the distribution (where even a small degree of skewness could drastically 
alter the results) suggests that any Gaussian analysis like the one above 
provides only a naive guess for the number of PBH's produced.  We must 
understand something about the true fluctuation statistics before making any 
assumptions about the nature of the probability distribution far from the mean.


\section{Stochastic Inflation}

The stochastic analysis of inflation~\cite{starb} 
provides an excellent method for obtaining statistics of scalar
field fluctuations.  In the stochastic approach, one divides the scalar
field driving inflation, $\p$, into a long wavelength
piece, $\p_{l}$, and a short wavelength piece, $\p_{s}$
\begin{equation}
  \p(x,t) = \p_{l}(x,t) + \p_{s}(x,t)
\end{equation}
where
\begin{eqnarray}
  \p_{l}(x,t) & = & \int d^{3}k \theta(\epsilon R(t)H - k) [a_{k}
	\varphi_{k}(t)e^{(-ik\cdot x)} + h.c.] \nonumber \\
  \p_{s}(x,t) & = & \int d^{3}k \theta(k - \epsilon R(t)H) [a_{k}
	\varphi_{k}(t)e^{(-ik\cdot x)} + h.c.]
\end{eqnarray}
and $\epsilon < 1$.
The long wavelength piece
contains only modes outside of the horizon: it is
coarse-grained over the horizon scale and will eventually 
be interpreted as the ``classical'' inflaton.  The short wavelength
piece includes only modes inside the horizon, where quantum fluctuations
are important.

The goal is to determine how the sub-horizon quantum fluctuations affect the 
evolution of the coarse-grained field.
We now briefly sketch the derivation of the stochastic slow-roll
equation. (For a more complete review see~\cite{yvbig} and
references therein.)  First, insert the decomposition (22) into
the Klein-Gordon equation, and write the result in the form
\begin{equation}
   \phidot_{l} + \frac{1}{3H}V'(\p_{l}) = f(\p_{s};x,t),
\end{equation}
where
\begin{equation} 
  f(\p_{s};x,t) = \frac{-1}{3H}\left[\phiddot_{s} + 3H\phidot_{s}
	-R^{-2}(t)\bigtriangledown^{2}\phi_{s}\right]. \nonumber
\end{equation}
As in the standard case (3), we have used the slow-roll approximation
on $\p_{l}$ and have neglected the spatial derivative since this piece
is homogeneous over the horizon scale.  The $\p_{s}$ piece 
has its dynamics dominated by the spatial derivative and any effects
of the potential are negligible.  The high frequency modes are then
approximated by those of a free massless field, $\varphi^{\star}_{k}(t)$,
governed by the wave equation
\begin{equation}
	(\frac{\partial^{2}}{\partial t^{2}} + 3H\frac{\partial}{\partial t}
		 + k^{2}e^{-2Ht})\varphi^{\star}_{k}(t) = 0,
\end{equation}
where we make the quasi-deSitter assumption: $H \approx constant$ and
$R(t) \sim$ exp$(Ht)$.
The solution to the free field equation is known
\begin{eqnarray}
 \varphi^{\star}_{k}(t)& = &\frac{H}{\sqrt{2k}}(\eta - i/k)\exp(-ik\eta) \\
 \eta & = &\int\frac{dt}{R(t)} \approx \frac{e^{-Ht}}{H}, \nonumber
\end{eqnarray}
so we have an explicit expression for $\p_{s}$ by letting
$\varphi_{k}(t) = \varphi^{\star}_{k}(t)$ in (23).  Inserting the expression
for $\p_{s}$ into
(25) we have
\begin{eqnarray}
  f(x,t) & = & \frac{-1}{3H}\left[\frac{\partial^{2}}{\partial t^{2}} 
  + 3H\frac{\partial}{\partial t}+ e^{-2Ht}\bigtriangledown^{2}\right]
  \int d^{3}k \theta(k - \epsilon R(t)H) [a_{k}
	\varphi_{k}^{\star}(t)e^{(-ik\cdot x)} + h.c.]  \nonumber \\
      & = & \frac{i \epsilon R(t) H^{2}}{\sqrt{2}(2\pi k)^{3/2}}
              \int d^{3}k \delta(k - \epsilon R(t)H)
              [a_{k}e^{-ik \cdot x} - h.c.]
\end{eqnarray}
and equation (24) is now complete.  The field $\p_{l}$ is
smoothed over the horizon scale and we are concerned with the
dynamics at a single spatial point.  Let $\p_{l}(x,t) \rightarrow
\p_{l}(t) \rightarrow \p(t)$ and interpret this 
piece as a classical field which is acted on by the stochastic force
$f(x,t) \rightarrow f(t)$.  The field $\phi$ is affected
not only by $V'$, as in the standard slow-roll approach, but
also by the flow of initially small-scale quantum fluctuations
across the horizon.  If we calculate the expectation
value and two-point function of $f(t)$ and interpret them
classically we obtain
\begin{eqnarray}
 \langle f(t) \rangle & = & 0, \nonumber \\
 \langle f(t)f(t') \rangle & = & \frac{H^{3}}{4\pi^{2}}\delta(t-t'),
\end{eqnarray}
and (24) becomes a Langevin equation for $\p$.
Regrouping terms and slightly changing notation, 
the final stochastic slow-roll equation is
\begin{equation}
 \phidot   =  - \frac{1}{3H(\p)}V'(\p) + \frac{H^{3/2}}{2\pi}g(t) 
\end{equation}
where $g(t)$ has the properties of Gaussian white noise,
\begin{eqnarray}
 \langle g(t) \rangle & = & 0, \nonumber \\
 \langle g(t)g(t') \rangle & = & \delta(t-t').
\end{eqnarray}
We have standard slow-roll evolution with an additional
stochastic term which represents the effect of quantum fluctuations
on the dynamics.

We will use the stochastic equation (29) to determine the probability
distribution of fluctuations, $P(\p,t)$, for models of interest, and
then measure the degree to which non-Gaussian statistics are important.
The Fokker-Planck equation associated with $P(\p,t)$ and (29) is
well known~\cite{risken}
\begin{eqnarray}
 \partial_{t}P(\p,t) = \frac{1}{3}\partial_{\p}
 \left[\frac{V'(\p)}{3H(\p)}P(\p,t)\right]   + \frac{1}{8\pi^{2}}\partial_{\p}
	\left[H^{3/2}(\p)\partial_{\p}\left(H^{3/2}(\p)P(\p,t)\right)\right],
\end{eqnarray}
with the Stratonovich interpretation of the noise.

Several authors have employed the stochastic Langevin equation and
corresponding Fokker-Planck equation to explore the importance of
non-Gaussian fluctuations for large-scale structure~\cite{stoch,yvbig,yvm}.  
We use a similar approach but on the much smaller scales relevant
for PBH formation.  We also choose to work with the more intuitive Langevin 
equation (29) when determining our solutions rather than the more complicated 
Fokker-Planck mathematics of (31).  Next we present several examples which 
illustrate our method of solution and provide a feel for
how non-Gaussian statistics arise from non-linear inflaton dynamics.

\bigskip

\noindent $\bullet$ \textbf{de Sitter: The Gaussian Case}

For our template example we examine the case of constant
vacuum energy, where there is no possibility of mode-mode 
coupling and the statistics should be exactly Gaussian.  
To show this point clearly, start with equation (29). We have
$H(\p) = H = constant$ and $V'(\p) = 0$, so the evolution is simple
diffusion
\begin{eqnarray}
 \phidot = \frac{H^{3/2}}{2\pi}g(t).
\end{eqnarray}
The normal statistical behavior becomes
apparent after integrating:
\begin{eqnarray}
 \p(t) = \p_{o} + \frac{H^{3/2}}{2\pi}\int_{t_{o}}^{t}g(t')dt'.
\end{eqnarray}
Since $\int_{t_{o}}^{t}g(t')dt'$ is Gaussian distributed as a noise
source, the resulting probability distribution for
$\p$ must be Gaussian as well.  From (33) we have 
\begin{eqnarray}
 \left<\p(t) - \p_{o}\right> & = & 0, \nonumber \\
 \left<\left|\p(t) - \p_{o}\right|^{2}\right> & = & 
	\frac{H^{3}}{4\pi^{2}}\int_{t_{o}}^{t}\int_{t_{o}}^{t}\delta(t'' -t')
	dt''dt' \nonumber \\
	& = & \frac{H^{3}}{4\pi^{2}}(t - t_{o}),
\end{eqnarray}
and the probability distribution  
\begin{eqnarray}
P(\p,t) & \propto & \exp\left[-\frac{(\p(t) - \p_{o})^{2}}
{2\sigma^{2}}\right]; \\   
\sigma^{2} & \equiv & \left<\left|\p(t) - \p_{o}\right|^{2}\right> 
        =  \frac{H^{3}}{4\pi}(t - t_{o}).\nonumber
\end{eqnarray}
Over a Hubble time ($t-t_{o} = H^{-1}$), 
$\sigma = H/2\pi$: the standard result from
quantum field theory.  

Non-Gaussian distributions arise when the 
evolution is non-linear, or when the inflation is not strictly de Sitter.  
One gains insight into the statistics of more
complicated examples by comparing them to the simple de Sitter case.
Our goal when
finding distributions for our examples will be
to put the stochastic equation into the form of (32) so that we
may simply write down the answer using (35). We illustrate this
technique in the examples that follow, and use the notation:
\begin{alignat}{2}
 x & \equiv \p/m_{pl}; & \qquad t &\rightarrow t/m_{pl}  \nonumber 
\end{alignat}
in order to keep the variables dimensionless.
\bigskip

\noindent $\bullet$  \textbf{Non-linear Diffusion: Driftless $\p^{4}$}

Mode-mode coupling arises when either the $V'$ ``drift'' term or the
diffusion term in (29) is non-linear in $\p$. In order to
 develop some intuition for how each of these effects generates non-Gaussian
statistics, consider first an example with non-linear diffusion when
the drift component is ignored ($V' \rightarrow 0$).
For concreteness let
\begin{eqnarray}
  V = \frac{\lambda}{4}m_{pl}^{4} x^{4}
\end{eqnarray}
 with
$H = m_{pl}\sqrt{\frac{2\pi\lambda}{3}}x^{2}$.  If we ignore
the drift term in (29) our stochastic equation is 
\begin{eqnarray}
\xdot = Cx^{3}g(t),
\end{eqnarray}
where $C = \lambda^{3/4}[54\pi]^{-1/4}$.
 The trick now is to change variables, $x \longrightarrow Z(x)$, such that
$Z(x)$ will satisfy the simple diffusion equation (32).  We want:
\begin{eqnarray}
 \Zdot = \frac{d Z}{d x} \xdot = \frac{d Z}{d x} 
    \left[Cx^{3}g(t)\right] = [constant]g(t),
\end{eqnarray}
 where we have used (37) in the second step.  Clearly if we let
\begin{eqnarray}
Z = \frac{1}{2}x^{-2}
\end{eqnarray}
then we are left with an equation of constant diffusion
\begin{eqnarray}
 \Zdot = Cg(t).
\end{eqnarray}
Now we write down the answer in correspondence with (35)
\begin{eqnarray}
P(Z,t) & \propto & \exp\left[-\frac{(Z - Z_{o})^{2}}{2C^{2}t}\right].
\end{eqnarray}
Finally, changing back to $x$ yields the probability distribution
of interest:
\begin{eqnarray}
P(x,t) = \left|\dZdx\right|P(Z(x),t) 
 	& \propto & x^{-3}
	\exp\left[-\frac{(x^{-2} - x_{o}^{-2})^{2}}{8C^{2}t}\right].
\end{eqnarray}

The statistics here are clearly non-Gaussian, but is the deviation
significant?  The answer depends on the value of $C^{2}t$
which is, in an approximate sense, the effective width of the
distribution, $\sigma_{eff}$. 
Expression (42) is plotted in Fig. 1
for $C^{2}t = 1$ and $x_{0} = 1$ with arbitrary normalization.  
We see that $P(x)$ has a long tail of large fluctuations,
and clearly is far from Gaussian.  For small
values of $\sigma_{eff}$, the deviation of $x$ from $x_{o}$ will not 
be significant
and the distribution will recover a Gaussian shape. 
We will discuss this behavior in detail following the next example.

\bigskip

\noindent $\bullet$ \textbf{Non-linear Diffusion: Full $\p^{4}$ Theory}

Now let us go one step further and include the diffusion term
for the potential (36).  We have $V' = \lambda m_{pl}^{3} x^{3}$ and
the stochastic equation (29) becomes
\begin{eqnarray}
   \xdot = -C_{1}x + C_{2}x^{3}g(t).
\end{eqnarray}
where $C_{1} = \sqrt{\frac{\lambda}{6\pi}}$ and $C_{2}
= \lambda^{3/4}[54\pi]^{-1/4}$.  
The drift term in this case is linear and does not lead to
mode-mode coupling.  The probability distribution of fluctuations
will then look very much like that of the driftless case (42), except
that the peak will shift to smaller $x$ values as the field
rolls down the hill, and the spread in the distribution with time
will no longer be linear.
 
As before, we obtain the solution
through a
change in variables: $x(t) \rightarrow Z(x,t)$.  In this case
we would like to first remove the drift term by effectively going to a
frame moving with the mean ``classical'' velocity (thus, the
explicit time dependence in Z).  By the classical
velocity we mean  the solution to (43) when the quantum fluctuating
is term left out:
\begin{eqnarray}
  \xdot = -C_{1}x,
\end{eqnarray}
which is simply
\begin{eqnarray}
  x_{cl}(t) = x_{o}\exp(-C_{1}t).
\end{eqnarray}
We would like our new variable to be constant as long as $x(t)$ follows
this classical path, so we let
\begin{eqnarray}
  Z(x,t) = x\exp(C_{1}t) 
        = \left(\frac{x}{x_{cl}(t)}\right)x_{o}.
\end{eqnarray}
We have chosen $Z(x=x_{cl}(t),t) = x_{o}$.  Now $Z$ has time derivative
\begin{eqnarray}
 \Zdot = \xdot \exp(C_{1}t) + C_{1}x\exp(C_{1}t),
\end{eqnarray}
and using (43) for $\xdot$ we have a stochastic equation for
$Z$
\begin{eqnarray}
\Zdot = C_{2}x^{3}\exp(C_{1}t)g(t) = C_{2}Z^{3}\exp(-2C_{1}t)g(t).
\end{eqnarray}   
Now our equation is much like (37) except for the extra time-dependent
factor.  We can, however, scale this factor out using a 
change in time variable: $t \rightarrow \tau(t)$.  If we remember that
$g^{2}(t) \sim \delta(t)$ then we have
\begin{eqnarray}
 \dZdtau = C_{2}Z^{3}\exp(-2C_{1})\left[\frac{dt}{d\tau}\right]^{-1/2}g(\tau).
\end{eqnarray}
In order to match (37) we need $d \tau/d t
 = \exp(-4C_{1}t)$, and 
requiring $\tau(t=0) = 0$ gives
\begin{eqnarray}
 \tau(t) = \frac{1}{4C_{1}}\left[1 - \exp(-4C_{1}t)\right] =
		\frac{1}{4C_{1}}\left[ 1 - (x_{cl}(t)/x_{o})^{4}\right].
\end{eqnarray}
We now have\footnote{From now on, 
the over-dot will signify the derivative with
respect to the argument of the stochastic function $g$.}
\begin{eqnarray}
 \Zdot = C_{2}Z^{3}g(\tau)
\end{eqnarray}
just as in (37),
and we simply use the solution (42) to give us the probability
distribution for $Z$
\begin{eqnarray}
 P(Z,\tau) \propto Z^{-3}\exp\left[-\frac{(Z^{-2} - Z_{o}^{-2})^{2}}
	{8C_{2}^{2}\tau}\right].
\end{eqnarray}
We transform back to $x$ and $t$ and obtain the standard result
for $\p^{4}$ theory~\cite{yvbig}
\begin{eqnarray}
  P(x,t) \propto x^{-3} \exp\left[-\frac{(x^{-2} - x_{cl}(t)^{-2})^{2}}
		{2\left(\frac{C_{2}^{2}}{C_{1}}\right)[\left(x_{o}
		/x_{cl}(t)\right)^{4}-1]}\right].
\end{eqnarray}
So, as expected, the distribution has the same non-Gaussian form
as did (42).  
In truth, however, most practical cases
($i.e.$ those concerning large scale structure) find this distribution 
quite well approximated by a Gaussian due to its extremely small
standard deviation, $\sigma \propto \sqrt{\lambda}$.  

To put in some numbers, let us assume we are interested in the 
statistics of fluctuations associated with
large-scales, say the modes that cross 
outside of the horizon during the epoch $x_{o} \sim 4$.
We need to evaluate (53) at $x_{cl}(t) =
x_{end} \sim 0.4$ in order to find the distribution of interest.
Since for this model 
$\lambda \sim 10^{-14}$, let us assume that the standard size of
a fluctuation about $x_{end}$ in the final distribution
is quite small: $x = x_{end} + \epsilon$ (where $\epsilon/x_{o}$ is assumed
$\ll 1$).   In (53) we have
\begin{eqnarray}
  x^{-2} \simeq x_{end}^{-2} - 2\epsilon/x_{end}^{3}
\end{eqnarray}
and
\begin{eqnarray}
 P(\epsilon) \propto \exp(\frac{-\epsilon^{2}}{2\sigma_{eff}}),
\end{eqnarray}
which is Gaussian as expected with an effective standard deviation
\begin{eqnarray}
  \sigma_{eff}^{2} =
	 \frac{x_{end}^{6}C_{2}^{2}}{2C_{1}}\left[(x_{o}/x_{end})^{4}
		- 1\right] = \frac{\lambda}{12}\left[(x_{o}/x_{end})^{4} -
		1 \right]x_{end}^{6}.
\end{eqnarray}
Plugging in the numbers we have $\sigma_{eff} \sim 2\times10^{-7}$.
So our approximation $\epsilon/x_{end} \ll 1$ is clearly self-consistent,
and the Gaussian assumption is a very good one 
in this typical case. 

The small value of $\sigma_{eff}$ is responsible for the near Gaussian
statistics, but the more the mean size of fluctuations increases
($\sigma_{eff} \uparrow$), the
more (54) fails and the Gaussian approximation (55) breaks down.  
In this way we can understand why a flat potential region
gives rise to non-Gaussianity.  Let us 
use
$C_{1}$ as a crude measure of the importance of the
drift term ($\sim V'$) in the dynamics (43).  
Notice that as the drift term gets smaller
($C_{1} \downarrow$), the effective width of the distribution (56) grows,
 and
with it the likelihood of non-Gaussian statistics.  Notice also that
as we consider fluctuations farther and farther from the mean, approximation
(54) will get worse and worse.  The number of standard deviations one must
be from the mean in order for non-Gaussian statistics to be important is
very sensitive to the value of $\lambda$.  However, we can at least
see the qualitative effect that large fluctuations have on highlighting
any intrinsic non-Gaussianity.

\bigskip
\noindent $\bullet$  \textbf{Non-linear Drift}

Finally we examine the case of non-linear drift as
a source of non-Gaussian behavior.  Consider the potential 
\begin{equation}
  V = \lambda m_{pl}^{4}(1 + ax^{3})
\end{equation}
over a region where $|x| \ll (1/a)^{1/3}$.  The
potential is roughly constant over this region 
($V \approx \lambda m_{pl}^{4}$) so the only non-linearity
comes from $V' = 3\lambda m_{pl}^{3} a x^{2}$.

  The stochastic equation in dimensionless variables is then
\begin{eqnarray}
 \xdot 
      & \approx  & -C_{1}x^{2} + C_{2}g(t),
\end{eqnarray}
where $C_{1} = a\sqrt{3\lambda/8 \pi}$, and $C_{2} = 
\lambda^{3/4}\left(32/27\pi\right)^{1/4}$.
Since the diffusion term is now constant, any non-Gaussian statistics will
arise due to the $x^{2}$ drift term in (58).  As before, we solve for the
probability distribution by making a change in variables,
$x(t) \longrightarrow Z(x,t)$.  We want
 to remove the drift term by
changing to a frame
moving with the classical velocity and canceling
out the drift in (58). The classical solution to the standard slow-roll 
equation of motion
\begin{eqnarray}
  \xdot = -C_{1}x^{2}
\end{eqnarray}
is trivial
\begin{eqnarray}
 x_{cl}(t) = (C_{1}t + x_{o}^{-1})^{-1},
\end{eqnarray}
where $x_{o} = x_{cl}(t=0)$.  Let us pick $x_{o} = -1$ for concreteness.
Our new variable, $Z(x,t)$, should be fixed as long as $x(t)$ travels
along its classical path:  $\frac{d}{dt}Z(x_{cl},t) = 0.$  We achieve
this goal by letting
\begin{eqnarray}
  Z(x,t) = (x^{-1} - C_{1}t)^{-1},
\end{eqnarray}
where we have picked $Z(x_{cl}(t),t) = x_{o}$. Differentiating
with respect to $t$ and using (58) we obtain a driftless
stochastic equation for our new variable
\begin{eqnarray}
 \Zdot 	= \frac{Z^{2}}{x^{2}} C_{2}g(t).
\end{eqnarray}
At this point let us simplify the problem by approximating $Z$ and
$x$ with their mean classical values: $x \rightarrow x_{cl}(t)$ and
$Z \rightarrow Z(x_{cl}(t),t) = x_{o} = -1$.  We have
\begin{eqnarray}
  \Zdot = (C_{1}t + 1)^{2}C_{2}g(t)
\end{eqnarray} 
which is beginning to look a lot like simple diffusion.  Clearly $Z(x,t)$
will be Gaussian distributed, but $\sigma(t)$ will be more complicated than
in (35) due to the explicit $t$-dependence in the diffusion coefficient.
To find the correct form of $\sigma(t)$ let us change time variables
$t \rightarrow \tau$ such that the time dependent diffusion will
be scaled away.  Remembering that $g^{2}(t) \sim \delta(t)$ we have
\begin{eqnarray}
	\dZdtau = \left[C_{1}t - 1\right]^{2}
		C_{2}\left[\frac{d\tau}{d t} \right]^{-1/2}
			g(\tau).
\end{eqnarray}  
So we need $d \tau/d t = (C_{1}t - 1)^{4}$, or
\begin{eqnarray}
  \tau = \frac{1}{5C_{1}}\left[1 + (C_{1}t - 1)^{5}\right].
\end{eqnarray}
Then our final stochastic equation for $Z(\tau)$ mirrors (32)
\begin{eqnarray}
  \dZdtau = C_{2}g(\tau),
\end{eqnarray}
and corresponds to the probability distribution (35):
\begin{eqnarray}
 P(Z,\tau) \propto \exp\left(-\frac{(Z + 1)^{2}}{2\sigma(\tau)^{2}}\right);
 \sigma(\tau)^{2} = C_{2}^{2}\tau. 
\end{eqnarray}
We obtain the distribution of interest, $P(x,t)$, by simply changing
back to the appropriate variables $Z \rightarrow x$, $\tau \rightarrow t$.
We have
\begin{eqnarray}
	P(x,t) \propto \left|\dZdx\right|\exp\left(-\frac{\left[(x^{-1}
		- C_{1}t)^{-1} + 1\right]^{2}}
	{2C_{2}^{2}\tau(t)}\right).
\end{eqnarray}
Note that the non-Gaussian behavior is evident not only in the non-linear
$x-$dependence of $Z$, but also in the non-linear time evolution of $\tau(t)$.
If we evaluate the distribution at a particular time,
say that corresponding
to $x_{cl} = -2$, we will have a peak in probability near this value.  
 From (44) we need $t = (2C_{1})^{-1}$, which corresponds to  $\tau \approx 
(5C_{1})^{-1}$ from (65), so (68) becomes
\begin{eqnarray}
	P(x) \propto \frac{1}{x^{2}}\exp\left[
	-\frac{\left([x^{-1} - 0.5]^{-1} + 1\right)^{2}}
	{2 \sigma^{2}}\right],
\end{eqnarray}
where $\sigma^{2} \approx C_{2}^{2}/5C_{1}$.

We plot (69) for $\sigma^{2} = 0.02$ in Fig. 2.  The non-Gaussian
behavior is clear.  There is a long tail in this distribution for small
fluctuations and a sharp cutoff for large fluctuations --- in opposition
to the previous example.  The non-linear drift tends to encourage
fluctuations to fall down the hill (since $V'' < 0$ the
slope gets steeper for smaller values of $x$) resulting in an under-abundance 
of large fluctuations compared to the Gaussian.   

\section{Toy Models}
In this section we present three example potentials which meet the 
small-scale power requirement for PBH production ($\delH \gsim 0.01$) 
as discussed in Sec. II.  
For each example we apply the methods developed in Sec. III to determine the 
probability distribution of fluctuations and compare this distribution
to the usual Gaussian assumption.  Each example is realistic in the
sense that it forces the power on large scales to be small 
($\delH \sim 3\times10^{-5}$) 
while giving us the sufficient small-scale power
needed.  The additional constraint on the inflaton potential is associated
with the over-production of gravitons and limits the scale
of inflation: $H/m_{pl} \lsim 10^{-5}$ during the epoch associated with
the horizon scale today ($\p(10^{4}$ Mpc$)$)~\cite{turner2}.  The first two
models below are actually on the hairy edge of this constraint, but recall
that we are not trying to construct ``true'' models here, we are simply 
interested in the non-Gaussianity associated with PBH production.  
Since the non-Gaussian statistics in these models derives from the 
\textit{shape} of the potential, slightly altering the scale of inflation
will not drastically affect the result.  The gravitational
wave constraint on our third toy model is a different story and
we will discuss the reason when we come to it.

We emphasize that these toy 
potentials are meant only to illustrate the importance of non-Gaussianity in
PBH producing models.  The correct distribution for any ``realistic'' model
will depend on the nature of the potential.  The idea is that
the following examples will be tools of pedagogy, enabling the reader to 
easily perform calculations of PBH abundances without relying
on an incorrect Gaussian assumption.
Before we begin, let us restate
\begin{eqnarray}
  x \equiv \p/m_{pl}, \quad \quad t \rightarrow t/m_{pl}, \nonumber
\end{eqnarray}
and introduce some notation
\begin{eqnarray}
 V = \lambda m_{pl}^{4} \Vtw, \quad \quad H = \sqrt{\lambda}m_{pl}/\Htw. 
   \nonumber
\end{eqnarray} 
The symbols $\Vtw$ and $\Htw$ are dimensionless quantities
of order unity which will be useful in keeping track of small parameters.

\subsection{Plateau Potential}
Consider the potential, shown in Fig. 3, which
 in some region of interest ($x \sim 0$)
has a flat ``plateau'' feature \footnote{This potential with a single
break is similar to the double-break potential proposed by Ivanov, Naselsky, 
and Novikov~\cite{ivan} for making PBHs.}
\begin{eqnarray}
  \Vtw = \left\{\begin{array}{ll}
		1 + \arctan(x) &\mbox{$x>0$} \\
		1 + (4\times10^{33})x^{21} &\mbox{$x<0.$}
		\end{array}
	\right. 
\end{eqnarray}
We have designed this somewhat outrageous potential to produce
a spike in small-scale power using the 
methods outlined in Sec. I.   It suffices to regard expression (70) as 
only a region of potential, so we arbitrarily begin following 
the roll-down of $x$ at the value corresponding to the scale $L
\approx 10$ pc, or $x_{st} \approx 0.15$.  We choose to normalize
our potential region at this hand-picked starting point, and 
fix $\lambda$ with $\delH(10$ pc$) = 3 \times 10^{-5}$, which is 
an arbitrary but reasonable choice.  
Expression (4) then requires $\lambda \approx 6\times10^{-10}$. 
When the field first reaches the plateau region, it is moving too fast 
to obey the slow-roll conditions, and we are
forced to obtain 
the form of $\delH$ by solving (1),(2) numerically.  
We also use the more exact
behavior $\delH \propto H^{2}/\xdot$.  The plateau region
is so flat, however, that the friction of the expansion
slows the field down very quickly.
The magnitude of $\delH$ grows 
as $x$ slows down and continues to grow until $\xdot$ reaches its lowest
value, corresponding to a peak in power, and also corresponding to the
beginning of a second slow-roll regime.
(Note that this potential provides two periods
of inflation with only one inflaton.)  Numerically, we find that
the peak in power
occurs when  
$x_{*}=-1.23\times10^{-2}$.  The height of the peak is then
\begin{eqnarray}
\delH(x_{*}) \sim \sqrt{\lambda}\frac{\Vtw^{3/2}}{\Vtw'}(x_{*}) \sim0.05.
\end{eqnarray}
We show the density perturbation spectrum at horizon crossing
associated with this region 
in Fig. 4, and see that the peak in power corresponds to 
the creation of $\sim 1M_{\odot}$ black holes.  

The modes which will eventually be responsible for
PBH production pass outside of the horizon at the epoch $x_{*}$.
Since we are only interested in calculating the non-Gaussian behavior
associated with these modes, we evolve the stochastic 
equation from $x_{*} \rightarrow x_{end}$ to determine the relevant
probability distribution.  For this model, the end of inflation occurs
when the first slow-roll condition breaks down: $\left|\frac{\Vtw''}
{24\pi\Vtw}\right| \approx 1,$ and corresponds to $x_{end}
= -1.55\times10^{-2}$.

The stochastic equation (29) in the new notation is
\begin{eqnarray}
  \xdot = \frac{-\sqrt{\lambda}\Vtw'}{3\Htw} + \frac{\lambda^{3/4}
	\Htw^{3/2}}{2\pi}g(t).
\end{eqnarray}
Note the relative sizes of the drift and diffusion terms in the
evolution of $x$. In most cases, the higher power of $\lambda$ in the
diffusion term means that quantum fluctuations play a very minor role
in the dynamics of $x$, resulting in nearly Gaussian statistics.
But in this example, the drift term ($\Vtw'$)
is very tiny over the region of interest, and the two terms
become of more equal weight, increasing the likelihood of non-Gaussian
statistics.

Over the range of interest, $|x| \sim 10^{-2}$, so to a high
degree of accuracy, $\Vtw \approx
1$, and we have a case of constant diffusion.
On the other hand, the drift force is small and very non-linear:
$\Vtw' = 8.4\times10^{34}x^{20}$.  This situation is
very much like the last example presented in Sec. III and the method 
of solution will be similar.  Let us re-scale $x$ by $x_{*}$
to keep our variables of order
unity: $\xtw = x/|x_{*}|$.  
The path of the mean evolution is then 
 from $\xtw_{*} =-1$ to $\xtw_{end} = -1.26$.
Using $\lambda = 6\times10^{-10}$ in (72)
we have our stochastic equation for $\xtw$
\begin{eqnarray}
  \xtwdot = -(1.2\times10^{-7})\xtw^{20} + (7.8\times10^{-6})g(t).
\end{eqnarray} 
We see that the small value of $\Vtw'$ gives the quantum diffusion term
a more equal weight in the dynamics.          
Let us do one more bit of cleaning up by rescaling the time variable
$t \rightarrow t(1.2\times10^{-7})$ (see (49)) which gives
\begin{eqnarray}
 \xtwdot = - \xtw^{20} + ag(t') \\
  a = 2.3\times10^{-2}.\nonumber
\end{eqnarray}
We now mirror the method of solution presented in Sec. III.  

The first step is to go to a frame moving with the classical velocity.
The classical slow-roll equation is $\xtwdot = \xtw^{20}$ and using
$\xtw_{o} = \xtw_{*} = -1$, the classical path is
\begin{eqnarray}
  \xtw_{cl}(t) = -[1 - 19t]^{-1/19}.
\end{eqnarray}
Our variable, $Z(\xtw,t)$, will remain constant along $x = x_{cl}$
\begin{eqnarray}
	Z(\xtw,t) = [19t - \xtw^{-19}]^{-1/19}
\end{eqnarray}
where we have chosen
$Z(x_{cl}(t),t) = -x_{o} = 1$.  The stochastic equation for $Z$ is
then diffusion-only by design
\begin{eqnarray}
 \Zdot = -\left(\frac{Z}{\xtw}\right)^{20}ag(t),
\end{eqnarray}
 and approximating $\xtw \rightarrow \xtw_{cl}(t)$, $Z \rightarrow -\xtw_{o}$
we have
\begin{eqnarray}
 \Zdot \simeq (1 - 19t)^{20/19}ag(t).
\end{eqnarray}
Now, with one more change of variables we can scale away the time
dependence multiplying g(t).  Let
\begin{eqnarray}
 \tau = \frac{1}{59}\left(1 - (1 - 19t)^{59/19}\right),
\end{eqnarray}
so that we are left with an equation exactly like (32)
\begin{eqnarray}
 \Zdot = ag(\tau).
\end{eqnarray}
The probability distribution, $P(Z,\tau)$, follows from (35), and we
change variables back again to obtain the distribution of interest:
\begin{eqnarray}
  P(\xtw,t) \propto \frac{1}{\xtw^{20}}\exp\left[
		\frac{-\left([19t - \xtw^{-19}]^{-1/19} - 1\right)^{2}}
		{2 a^{2} \tau(t)}\right].
\end{eqnarray}
We want to evaluate the distribution at $t_{end}$, the time
corresponding to $\xtw_{cl}
= -1.26$.  From (75) we find $19t_{end} = 0.9876$ and from (79)
we have $\tau(t_{end}) \simeq 1/59$.  Plugging in these 
values along with $a = 2.3\times 10^{-2}$, we have the final distribution
of fluctuations:
\begin{eqnarray}
  P(\xtw) \propto \frac{1}{\xtw^{20}}\exp\left[-K\left(
	\left[19t_{end} - \xtw^{-19}\right]^{-1/19} -1\right)^{2}\right], 
\end{eqnarray}
where $K = 5.6 \times 10^{4}$. 
The distribution is plotted in Fig. 5 along with the results
of a numerical simulation for this same potential.  For our simulation,
 we started with
the stochastic equation (72) and made no approximations.  We used 
the Box-Muller method~\cite{boxmuller} to 
transform uniform deviates into random Gaussian
deviates to mimic the stochastic force.  The numerical results
consist of 
$4\times10^{4}$ individual runs of the Langevin equation, and we have
normalized the height of our calculated distribution to fit this number.
We see that the calculated distribution (82) agrees well with the 
numerical results.  

The distribution of fluctuations is clearly non-Gaussian.  In Fig.
6 we have plotted our analytic distribution (82) along with two
Gaussians that one may wish to compare it to: one with the same
mean and standard deviation as our distribution, and one with
the same height and width at half maximum as the peak in our distribution.
We see that the true distribution is under-producing large fluctuations
compared with either of the Gaussians.  Now, for PBH production under
the Gaussian assumption, we are concerned with fluctuations on the
tail of the distribution $\sim
6\sigma$.  The calculated distribution differs so drastically from the Gaussian
assumption at this distance from the mean, that we can only compare them
on a log scale.  Fig. 7 shows distribution (82) along with the 
two Gaussian comparisons in units of the standard deviation from the
mean.  As the more conservative choice, we use the $\sigma$ associated 
with the Gaussian 
fitted to the peak of (82).  Observe that the Gaussian
assumption would vastly
over-produce PBHs, with an error of order $\sim 10^{150}$ at $6\sigma$!  
This model was designed to
give us a large number of PBH's using the Gaussian assumption, but
upon examining the distribution more closely, we see that the actual
production is practically $zero$.  We, as inflation designers, are forced
to make an even higher spike in small-scale power if we want PBH production.
Note also that as 
the spike becomes higher, the drift term becomes smaller, and the 
distribution will tend to skew
even more towards small fluctuations.

\subsection{Wiggle Potential}

Our second potential which produces significant small-scale power
is one with a wiggle in the path of the inflaton 
\begin{eqnarray}
\Vtw = 1 + [136.717]x^{3} - 0.05x
\end{eqnarray} 
as shown in Fig. 8.  We start our evolution at $x(L = 10^{4}\quad{\rm Mpc}) = 
2.6$, and since slow-roll is obviously invalid over the dip,
we integrate (1),(2) to obtain the spectrum of $\delH(L)$ shown in Fig.
9.  Normalizing at $x = 2.6$ to COBE gives 
$\lambda \sim 5\times10^{-14}$. 
We have adjusted the wiggle to make the field
slow down dramatically just after the top of the bump ($x_{top}
=-1.0411\times10^{-2}$) which produces 
a large spike in power corresponding to the
$\sim 10^{28}$ g mass scale. We find the slow-point to be $x_{*}
= -1.10448\times10^{-2}$, which is also the beginning
of a second slow-roll epoch of inflation.  The height of the peak is
$\delH \sim \sqrt{\lambda}\frac{\Vtw^{3/2}}{\Vtw'}(x_{*}) 
\approx 0.01$,
the order of magnitude we need for PBH production.       

After the field slows down at $x_{*}$, inflation continues until $x_{end}
\sim -0.1$.  This is the path of interest for estimating 
the distribution of fluctuations.
Let us again set our notation before writing down the Langevin
equation.  For the analytic calculation we use $\Vtw \approx 1$ over
the range of interest\footnote{This approximation is very good at $x_{*}$ but
is off by $\sim 10\%$ at the end of inflation.  However, since
the field spends 
most of its evolution time near $x_{*}$, this approximation should be fine.}
and also scale $x$ by $\xtw = x/x_{top}$, 
such
that $\xtw_{top} = -1$ corresponds to the top of the wiggle 
($\Vtw'(\xtw = -1) = 0$).  Using this notation, the starting point
is $\xtw_{*} = -1.00033$ and we want to follow the evolution 
until $\xtw_{end} \sim -9$.
The potential in terms of $\xtw$ is
\begin{eqnarray}
 \Vtw  \approx  1, \quad \quad \Vtw' =  0.05[\xtw^{2} -1] \nonumber 
\end{eqnarray}
and the stochastic equation for $\xtw$ follows
\begin{eqnarray}
 \xtwdot = -(1.2\times10^{-7})(\xtw^{2} -1) + (7.5\times10^{-9})g(t).
\end{eqnarray}
As before, we clean things up with a new time scale 
$t \rightarrow t(1.2\times10^{-7})$, which gives us
\begin{eqnarray}
 \xtwdot = -(\xtw^{2} - 1) + ag(t), \quad
 a = 2.2\times 10^{-5}. 
\end{eqnarray}
We want to use the same trick and factor out the drift 
term by changing to a variable
which is constant along the classical path.  We solve
$\xtwdot = -(\xtw^{2} -1)$ to obtain the classical solution with
the initial value $\xtw_{cl}(t=0) = \xtw_{*} = -1.00033$ and obtain
\begin{eqnarray}
 \xtw_{cl}(t) = \coth(t - 4.35).
\end{eqnarray}
Following the methods outlined in Sec. III, the new variable, 
$Z(\xtw,t)$, is
\begin{eqnarray}
	Z(\xtw,t) & = & \coth[\coth^{-1}(\xtw) - t]
\end{eqnarray}
where
\begin{eqnarray}
  Z(\xtw_{cl}(t),t) = \xtw_{*}.
\end{eqnarray}
The stochastic equation for $Z$ is then
\begin{eqnarray}
 \Zdot = a\frac{Z^{2} -1}{\xtw^{2} - 1}g(t).
\end{eqnarray}
If we use the approximation $\xtw \rightarrow \xtw_{cl}(t)$,
$Z \rightarrow \xtw_{*}$ we have
\begin{equation}
\begin{split}
  \Zdot &=  \frac{a(\xtw_{*}^{2} - 1)}{\coth^{2}[t - 4.35] - 1}g(t) 
			\nonumber \\
	\nonumber \\
	&=  \sinh^{2}[t - 4.35]bg(t), \\
\end{split}
\end{equation}
where $b = 1.47\times10^{-8}$.  Now we want to change to
a new time coordinate, $\tau(t)$, which will leave us with a simple diffusion
equation.  The requirement is
\begin{eqnarray}
   \frac{d\tau}{dt} = \sinh^{4}[t - 4.35],
\end{eqnarray}
or
\begin{eqnarray}
 \tau(t) = C + \frac{3}{8}[t-4.35] -\frac{1}{4}\sinh[2(t-4.35)] 
		+\frac{1}{32}\sinh[4(t-4.35)],
\end{eqnarray}
where $C = 5.6\times 10^{5}$ demands that $\tau(t = 0) = 0$.  So the 
variable $Z(\xtw,\tau)$ obeys simple diffusion by design
\begin{eqnarray}
 \Zdot = bg(\tau)
\end{eqnarray}
and has a Gaussian probability distribution with mean $\bar{Z} = \xtw_{*}$
and $\sigma^{2} = b^{2}\tau$.  
Now, as discussed earlier, we simply perform the reverse transform
$Z \rightarrow \xtw$, $\tau \rightarrow
t$ to obtain the distribution of inflaton fluctuations:
\begin{eqnarray}
P(\xtw,t) \propto \left(\frac{1}{\xtw^{2} - 1}\right)
		 \exp\left[-\frac{(\coth(\coth^{-1}(\xtw) - t) - \xtw_{*})^{2}}
		{2b^{2}\tau(t)}\right].
\end{eqnarray}
We want to evaluate the above expression at the time $t_{end} = 4.24$,
when the classical path
reaches $\xtw_{end} = -9$.  Equation (91) gives us $\tau(t_{end})
\approx C$ and plugging in all of these values we have the final
probability distribution
\begin{eqnarray}
  P(\xtw) \propto \left(\frac{1}{\xtw^{2} - 1}\right)\exp\left[-K
	\left(\coth(\coth^{-1}(\xtw) - 4.24) + 1.00033\right)^{2}\right],
\end{eqnarray}
with $K = 4.1\times10^{9}$.
As in the previous example, we have simulated this distribution
numerically using no approximations, and we plot the two together in Fig.
10.  We have normalized (94) to the simulation height.  We see again that
the derivation does quite well, and the distribution has
a deficit of large fluctuations relative to a Gaussian.  The distribution
is skewed negative again since the drift term $\sim V'$ and
$V'' < 0$.  Negative fluctuations tend
to fall down the hill more quickly than positive ones.  There is no need
to do another explicit comparison to a Gaussian distribution since if
the distribution is clearly non-Gaussian to the eye, then the high-$
\sigma$ tail will be exponentially worse.

\subsection{Cliff Potential}
For our last example we present a potential region which flattens
out to achieve large power in the form of a blue spectrum, and then 
has a cliff-like feature where the slope abruptly becomes much steeper.   
Consider the region of potential, shown in Figure 11, given by
\begin{eqnarray}
  V = \lambda m_{pl}^{4}\left\{\begin{array}{ll}
		\cos^{-2}[1.5x]  &\mbox{$x>0.00004$} \\
	       (2.7)^{-4}[x + 2.7]^{4}  &\mbox{$x<0.00004$}.
		\end{array}
	\right. 
\end{eqnarray}
For $x>0.00004$ we use a form suggested by Hodges and Blumenthal
~\cite{hodges} which gives us $\delH
\sim \left(L/L_{o}\right)^{-n}$,
$n\simeq 0.18$.  The highest amplitude in power, 
corresponding to the flattest region of the potential and 
PBH formation, occurs
at $x_{*} = 0.00004$, after which the slope increases abruptly as a
`` $\p^{4}$ '' form ends inflation.  
Since we are interested in fluctuation statistics
which depend only on the nature of the potential from
$x_{*} \rightarrow x_{end}$, the ``small $V'$'' argument does not apply  
to encourage non-Gaussian statistics in this case.      
Using $\lambda \sim 3\times10^{-11}$ we obtain the power spectrum
shown in Fig. 12.  The highest point in density is $\delH 
\sim 0.04$ and corresponds to $\sim 10^{20}$ g PBH's.

As we mentioned before, the region of interest for calculating
the probability distribution has quartic form. Recall that we
have already calculated the distribution for the quartic case in 
Sec. III (53) so our job is nearly done.
In order to make things look more familiar,
define $y \equiv x + 2.7$, $\lambda' = 4\lambda/(2.7)^{4} \sim 2.
\times 10^{-12}$, such
that we have the more standard form
\begin{eqnarray}
 V(y) = \frac{\lambda'}{4}y^{4},
\end{eqnarray}
 and we are interested in the path from $y \approx 2.7$ to $y_{end} \approx
 0.4$.
Now, letting $\lambda \rightarrow \lambda'$ and $x \rightarrow
y$ in (53) we have the fluctuation distribution we need
\begin{eqnarray}
 P(y) \propto y^{-3}\exp\left[-\frac{(y^{-2} - y_{end}^{-2})^{2}}
		{2\sigma^{2}}\right], \\
	\sigma^{2} = \frac{\lambda '}{3}[(y_{*}/y_{end})^{4} - 1]. \nonumber 
\end{eqnarray}
 From the discussion following (53), we know that P(y) is skewed
positive, but we also know that the amount of skewness depends on the
value of $\sigma^{2}$.  In this case, non-Gaussianity is negligible due
to the extremely tiny value of $\lambda'$.
In order to see the lack of skewness in a quantitative way,
let us borrow a measure proposed by Yi, Vishniac, 
and Mineshige ~\cite{yvm}.  This simple estimate for skewness is the ratio
\begin{eqnarray}
 R \equiv \frac{P(y_{end} + N\sigma_{eff})}{P(y_{end} - N\sigma_{eff})}
	& \approx & \exp\left[N^{3}\sqrt{3\lambda'/4}
	(y^{4}_{*} - y^{4}_{end})^{1/2}\right] \approx \exp\left[
		N^{3}(10^{-5})\right]
\end{eqnarray}
where $\sigma_{eff}$, the number of effective standard deviations from the
mean, was given in (56), and $N$ measures the size of the fluctuation.
We see clearly that the small value of $\lambda'$ tends to
force $R \rightarrow 1$ and prevent skewness.  But a large value of 
$N$ (\textit{e.g.} a large fluctuation) will
compete with this effect and give a skewing effect.
For our potential (94) we have chosen the largest value of $\lambda \rightarrow
\lambda'$ possible to be consistent with the gravitational wave constraint
$\Htw \sim 10^{-5}$.  However, even with the large fluctuation
we need for PBH formation, $N\sim10$, expression (97) implies that the skew 
is only $1\%$.  Suppose, for example, that we did ignore the gravity wave 
constraint~\footnote{Note
that we can easily adjust our potential region to obtain the appropriate 
small scale power with a new value of $\lambda$.}
and push the value of $\lambda'$ up to $\sim 10^{-9}$.  The ratio (97)
would then give us a $\sim 20\%$ effect at $N=10$.  Thus
even for potentials with ``normal'' shapes, non-Gaussianity
may be important due to the rare large fluctuations associated with
PBH production.

\section{Conclusion} \label{sec:conclusion}
We have argued that the very conditions associated with PBH formation:
\begin{enumerate}
\item Large spikes in power, usually associated with flat potential regions
\item Rare, many-$\sigma$ fluctuations
\end{enumerate}
are exactly the same conditions which can produce
significant non-Gaussian fluctuations.  Flat potential regions promote
the importance of quantum fluctuations in the inflaton dynamics and
encourage the mode-mode coupling responsible for non-Gaussian statistics.
On top of this effect, many-$\sigma$ fluctuations push us out to the
tail of the probability distribution, where any intrinsic skewness 
will be amplified.  We have quantified this intuition with several toy 
models that produce the small-scale power associated with PBH production, 
and have used the the stochastic slow-roll equation to obtain the fluctuation 
distributions.  Our examples clearly illustrate that the Gaussian assumption 
can lead to large errors in the calculated number density of PBH's, and that 
the nature of the non-Gaussian distribution is extremely model dependent.  

Specifically, for models with spikes in small scale power, the fluctuation
distributions were skewed towards small fluctuations, under-producing
PBH's by many orders of magnitude relative to the Gaussian assumption. 
The negative skewing in these examples came from mode-mode coupling
due mainly to non-linear drift, which encouraged negative 
fluctuations to fall down the hill faster than positive ones 
($V'' < 0$)~\footnote{A potential region which tends to ``cup'' 
the inflaton ($V'' > 0$) after the flat region may produce positive skewing.}.
Because the fluctuation statistics depend on  
the path of the inflaton \textit{after} it passes the
flattest region of the potential, we were even able to construct an example 
where the Gaussian assumption holds, simply by forcing a dramatic
increase in  $V'$ (a ``cliff'') just after the peak in power.  
One may regard such a cliff region as unnatural, 
but it does illustrate the model-dependent nature of distribution shapes.
    
These results have several important implications
and we discuss each briefly.
First, the standard approach for limiting the initial fraction of PBH's, 
$\beta$, and limiting the spectrum of initial density perturbations 
(see Sec. II) must be reconsidered.  Because of the model-dependent nature
of the distributions, it is not possible, as in the standard 
practice~\cite{carrlid}, to 
use PBH over-production to limit generic inflationary power 
spectra.  Correspondingly, it is not possible to determine the 
number of PBH's produced from the power spectrum alone.  We
can use our toy models to exemplify what might happen
in various cases.   For models with spikes in power, like our plateau
and wiggle examples, fluctuation 
distributions will probably skew towards small fluctuations and
under-produce large fluctuations relative to the Gaussian case.
Many-$\sigma$ fluctuations then will be much less likely, 
and PBH production in these models will require an even higher
spike in power.  Then magnitude limits on spikes
in power to prevent PBH over-production will be less stringent 
than the limits obtained using Gaussian statistics.  We also point out that 
the formation of any PBH's at all without over-producing them  
will require an even more drastic fine-tuning than in 
the Gaussian case.  The fine-tuning must be more precise because the 
large fluctuation tails in these models are much steeper than a Gaussian tail. 
So small changes in fluctuation size will cause a larger change 
in the probability of having such fluctuations.  

Our results also affect the use of PBH over-production
to rule out or limit the parameter space of specific inflationary models.
For example, authors employ PBH over-production to constrain
the slope of blue perturbation spectra from inflation (see~\cite{carrnew}).  
All such examinations use the Gaussian assumption, and limit models
without regard to the shape of the potential after the power reaches
its maximum height.  But the amount of
PBH production, and hence constraints on $\delH$ and the slope
of the spectrum, depend crucially
on the nature of the fluctuation distribution and hence on the
shape of the potential between the region of PBH formation and
the end of inflation.  Our ``cliff'' potential
toy model actually recovers the Gaussian approximation, but only because
the slope of the potential increases dramatically just after the region
of high power associated with PBH production.  However, with a 
more rounded potential shape after the peak instead of a cliff, 
non-Gaussian fluctuations will be much more likely.  
PBH over-production is also important for
constraining parameter space in hybrid inflationary 
models due to associated spikes in small-scale power~\cite{guth2,linde2}.
Again, our toy models indicate that spikes in power are associated
with PBH under-production relative to the Gaussian case.  However, without 
further investigation of multiple-field PBH production models, the
validity of our intuition from single-field examples remains 
unclear~\footnote{An analytical examination
of multiple field models may prove to be too difficult, but numerical
simulation is always possible.}.

Authors occasionally investigate the possibility of PBH formation
associated with a soft equation of state during a $p=0$ ``dust like''
epoch~\cite{poln,carrnew}.  Dust era PBH formation would be important 
if the universe underwent an early $p=0$ stage (before the usual 
matter-dominated epoch).  Conceivably, such a dust stage could occur 
due to some as yet unknown physics (\textit{e.g.} possibly due 
to coherent inflaton oscillation during reheating~\cite{klop}). 
For dust era PBH formation, the
major criterion for PBH collapse is that the initial perturbations be 
non-rotating and spherically symmetric~\cite{poln,carrnew}, and the 
probability for spherical geometry typically increases with the amplitude
of a fluctuation.  Again the associated power
spectra must have excess power on small scales to achieve
substantial PBH production, and the fluctuation distributions will typically
be non-Gaussian.  But the tails of probability 
distributions are not so important to the analysis since rare 
large-amplitude fluctuations are not all that is important here, 
but also spherical ones.  Thus the value of $\beta$ 
will depend mainly on (the rms) $\delH$, with only a weak dependence
on the shape of the distribution.  Compared to PBH formation 
during the radiation dominated era (where, as we have shown, the distribution
is very important and non-Gaussianity can shift $\beta$ by many 
orders of magnitude), the effect of non-Gaussianity on dust era formation 
should be much smaller.  
  
Finally, we discuss our results in the context of PBH formation associated
with the QCD phase transition.  Jedamzik~\cite{jed} points out that 
PBH formation due to a first order QCD phase transition
would be roughly consistent with $\sim 0.5 M_{\odot}$ MACHO's,
since this is roughly the total mass-energy inside the horizon during the
QCD epoch.  However, even if the QCD transition is first order, the minimum  
$\delH$ for PBH formation at this mass scale will be lessened by at 
most a factor of order unity, for the following reason.   During the epoch of 
the first order phase transition, the effective velocity of sound drops to 
zero, and with it the Jeans mass.  Thus density perturbations which cross 
inside the horizon at the beginning of the epoch can begin to grow.  
But the duration of the phase transition is quite short.  Fluctuation 
amplitudes on this scale will be enhanced slightly relative to the 
standard case before the universe again achieves a hard equation of state.  
If the fluctuation amplitude at this time
is as large as $\delta \sim 1/3$ then PBH formation can occur.  So 
the minimum value of the initial $\delH$ at horizon 
crossing is reduced, but only by a factor of order unity 
(see~\cite{schmid} for a more complete
discussion of perturbation growth during this epoch).
This slight increase in the PBH mass function would be important
if PBH production were marginally possible over some range of
mass scales.  PBH formation would then be enhanced for
masses around $\sim 0.5M_{\odot}$, which could perhaps explain why this 
mass range of MACHO's is observed.  But in order to achieve PBH formation
at the QCD mass scale, we still must have large-amplitude fluctuations
at this wavelength, so our arguments for non-Gaussianity still apply.  
Moreover, any non-Gaussianity in the fluctuation distribution should be 
important, since PBH formation is again associated with rare fluctuations, 
and therefore quite dependent on the shape of the distribution far from
the mean.

\bigskip

\acknowledgments
JSB acknowledges support from a GAANN predoctoral fellowship at UCSC,
and JRP is supported by NSF and NASA grants at UCSC.

\bigskip
\bigskip

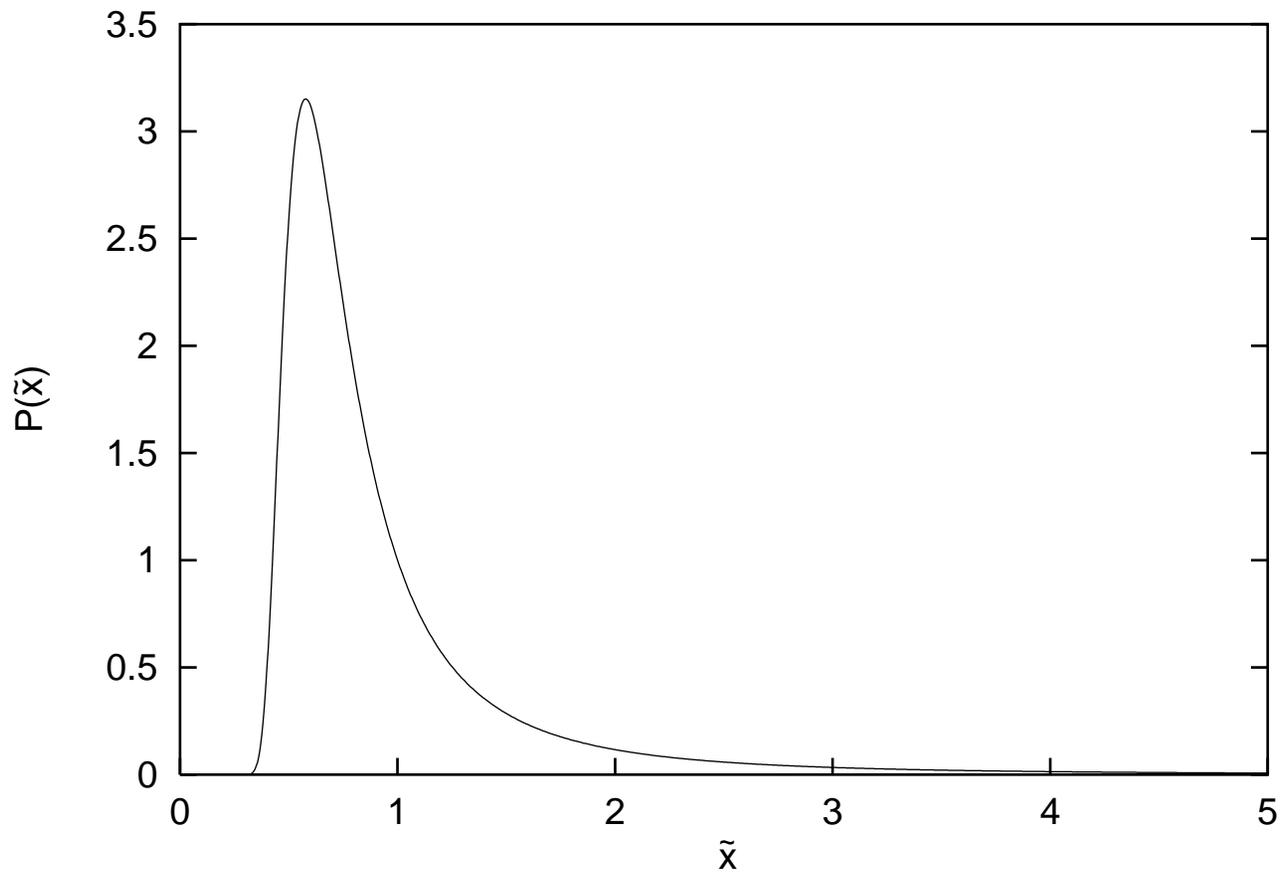
\begin{figure}
\caption{ The unnormalized probability distribution (42) 
for the driftless $\p^{4}$ example, where $x \equiv \p/m_{pl}$.  We use the
the value $x_{o} = 1$ arbitrarily and pick $C^{2}t = 1$, which is
unnaturally large, in order to demonstrate the intrinsic non-Gaussianity
resulting from non-linear diffusion.}
\label{fig1}
\end{figure}

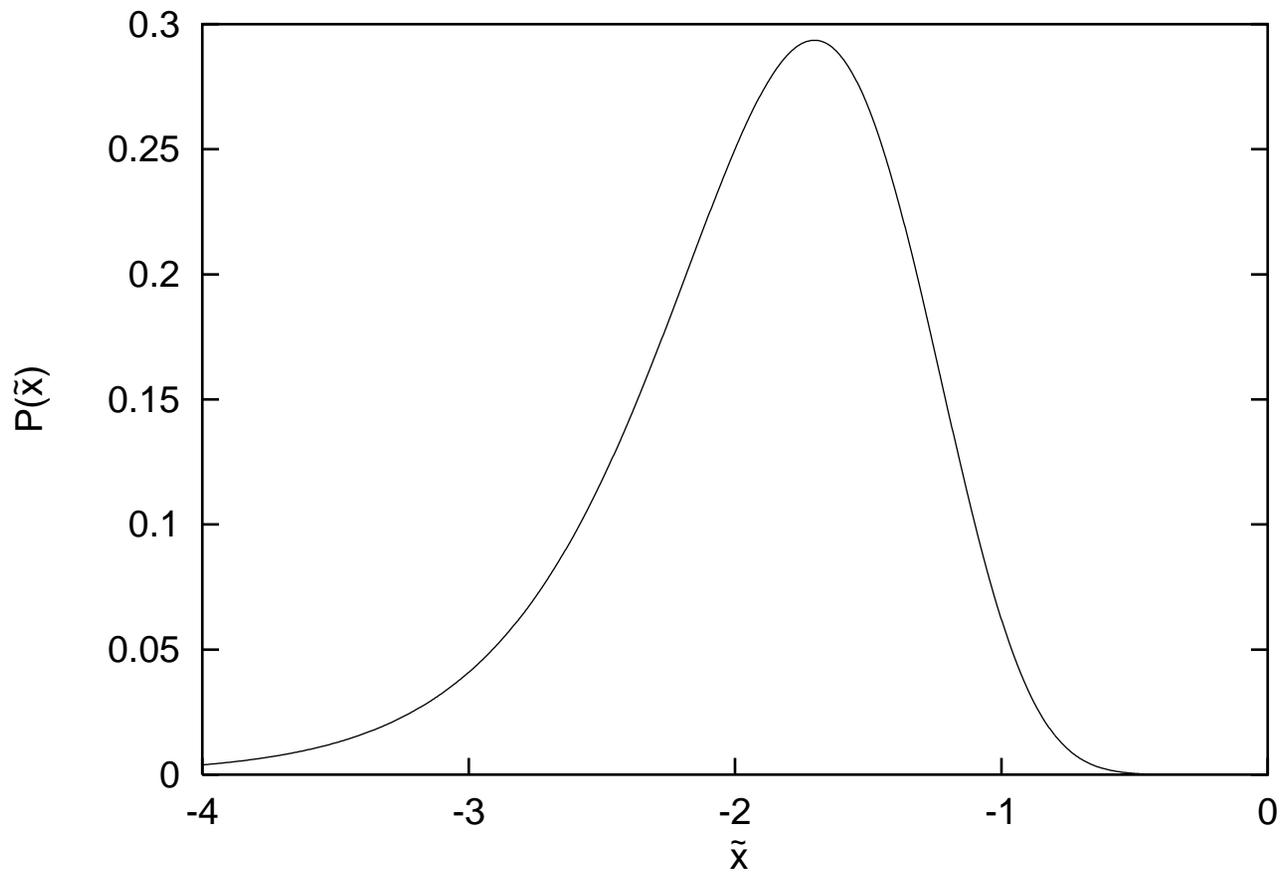
\begin{figure}
\caption{The unnormalized probability distribution (69)
for the non-linear drift example.  We have used $\sigma^{2} = 0.02$.
Note that the non-linear drift has caused negative skewing, in contrast
to Fig. 1, due to the tendency for negative fluctuations to fall down the hill
more quickly than positive ones.}
\label{fig2}
\end{figure}

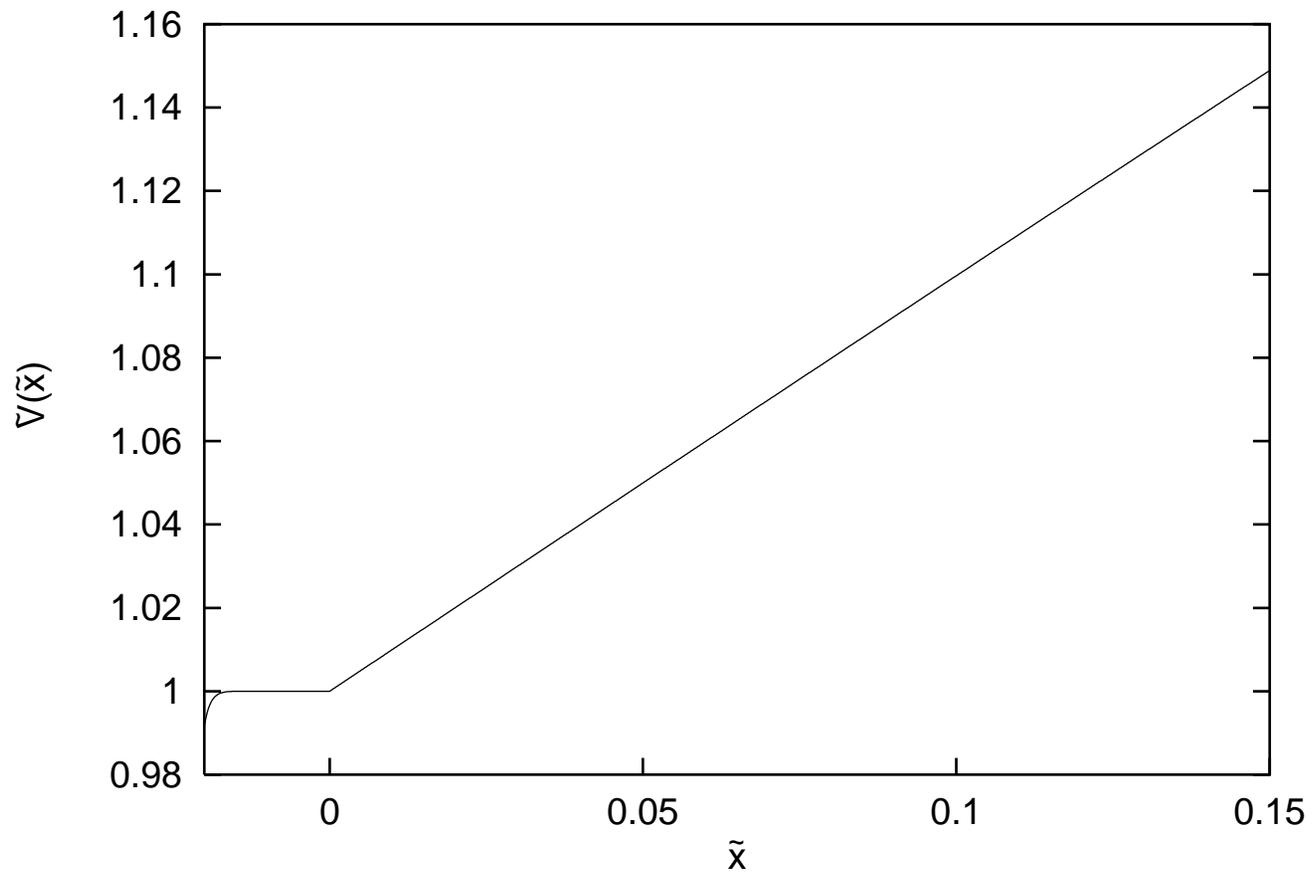
\begin{figure}
\caption{ The plateau potential (70): $\tilde{V}(x > 0) =
1 + \arctan(x)$, $\tilde{V}(x < 0) = 1 + 4\times10^{33}x^{21}$, where
$\tilde{V}$ is defined to be dimensionless and of order unity, $\tilde{V}
\equiv V/(\lambda m_{pl}^{4})$.  This potential
produces the fluctuation spectrum shown in Fig. 4 and the distribution
of fluctuations shown in Fig. 5.}
\label{fig3}
\end{figure}

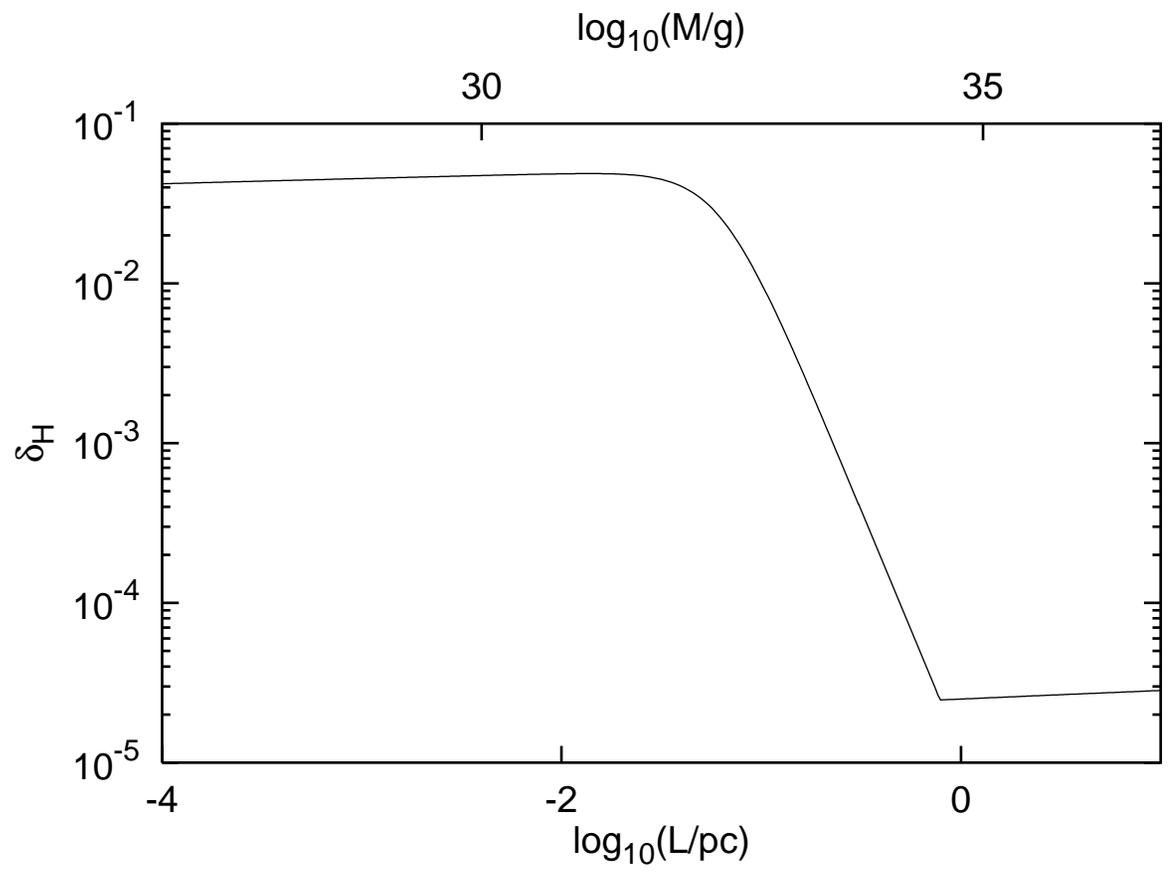
\begin{figure}
\caption{The spectrum of density fluctuations at horizon crossing, 
 $\delH$, associated with the plateau model (70) as a function of
the logarithm of the length scale $L$ in units of pc.  The plateau 
region seen in Fig. 3 produces the rapid rise in power, corresponding to
$\sim 10^{32}$g PBH production.}
\label{fig4}
\end{figure}

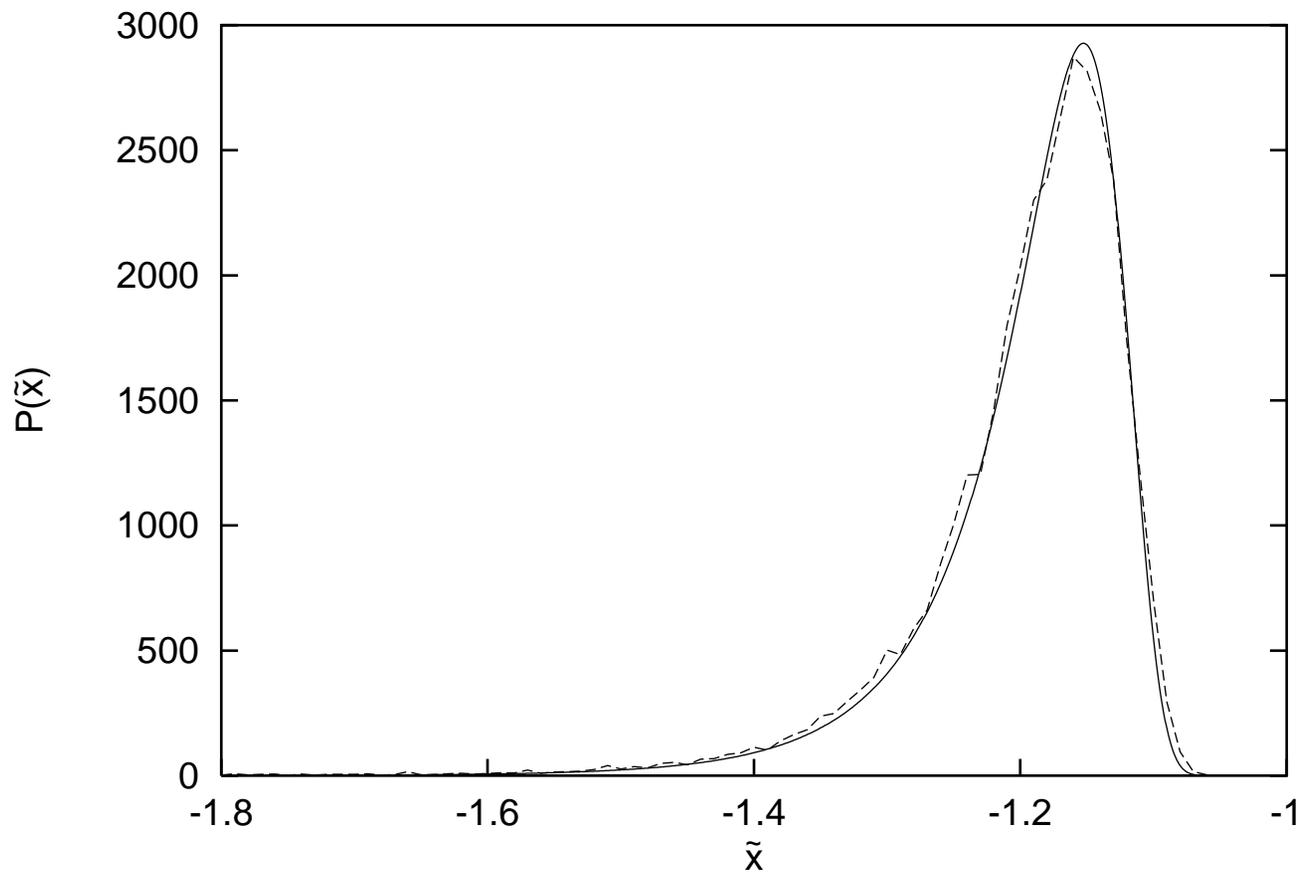
\begin{figure}
\caption{ The solid line is the calculated 
final distribution  of fluctuations (82) associated with PBH 
production from the plateau 
potential (70), Fig. 3.  The dashed line is a stochastic numerical 
simulation of the the same model which consists
of $4\times10^{4}$ points.  The calculated distribution is normalized to
to the number of points and bin size of the numerical result. }
\label{fig5}
\end{figure} 

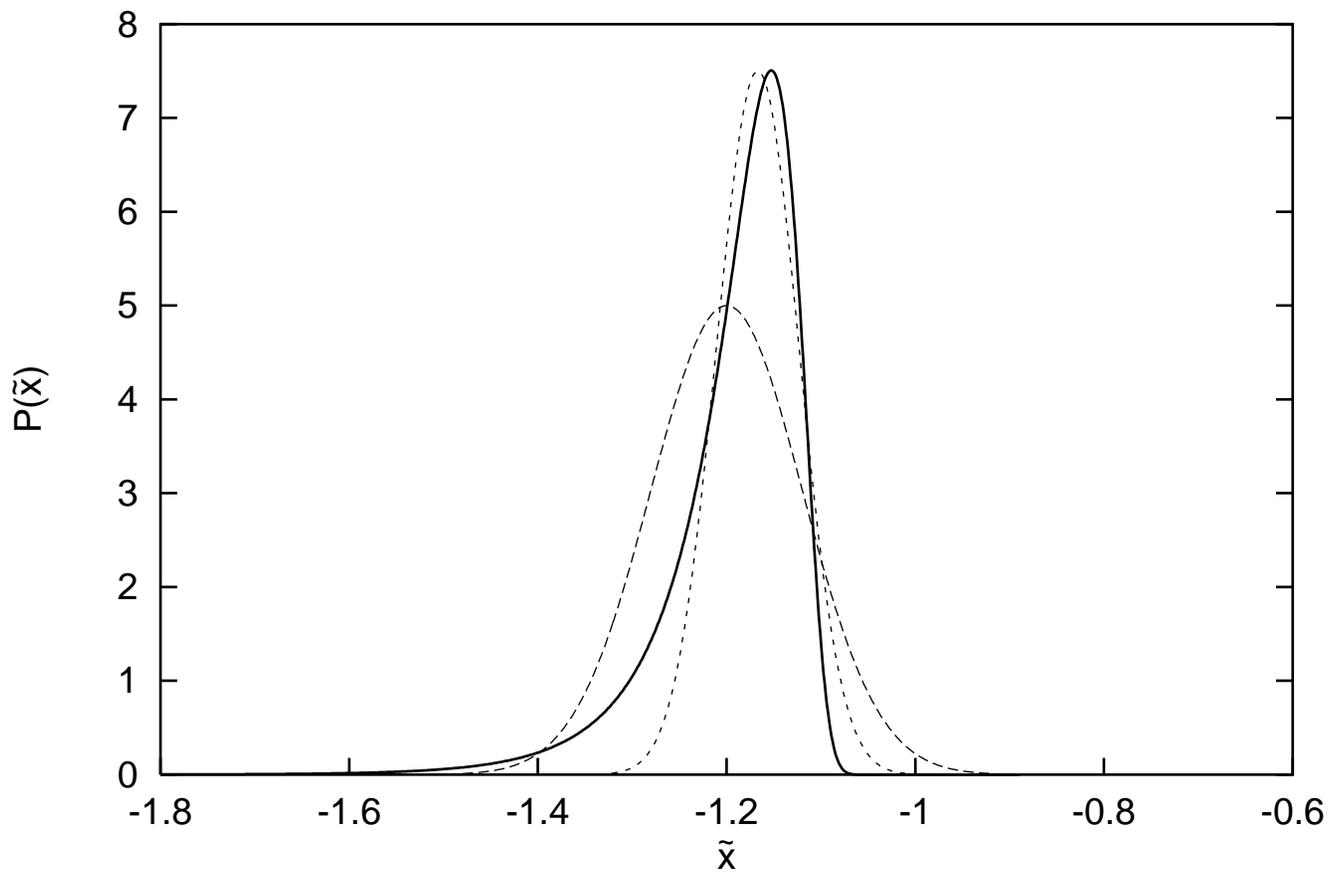
\begin{figure}
\caption{A comparison of the calculated distribution
of fluctuations for the plateau potential (70) with two Gaussian
counterparts.  The solid line is the analytic result (82).  The dashed
line is a Gaussian with the same mean and standard deviation as the
calculated distribution.  The short-dashed (dotted) line is a Gaussian 
with the same width at half maximum as the peak of the calculated 
distribution.  Both Gaussians over-populate large fluctuations.}
\label{fig6}
\end{figure}

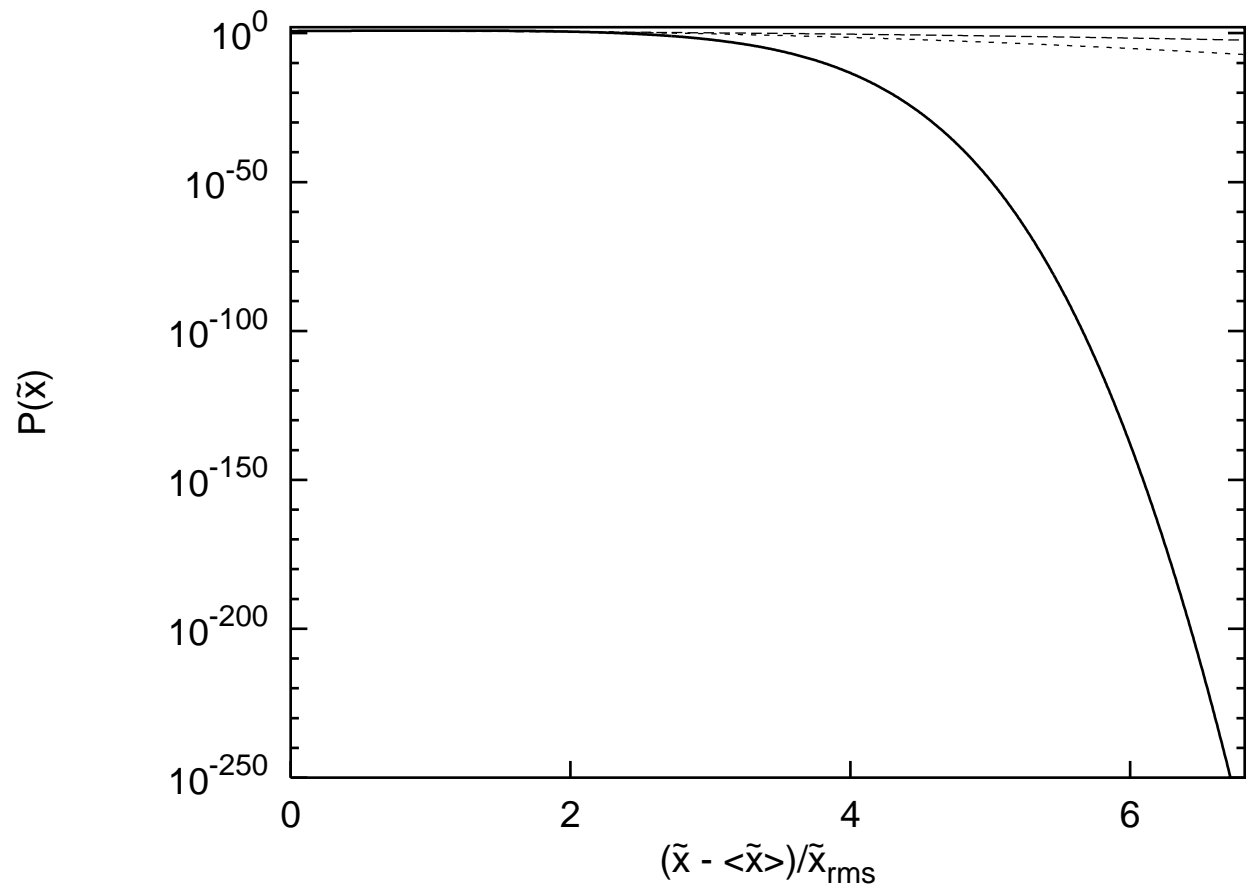
\begin{figure}
\caption{The same curves as in Fig. 6, but on a
log scale to emphasize the error in the Gaussian assumption.  Again,
the solid line is the probability distribution (82) associated with 
the plateau potential (70).  The two dashed lines are possible Gaussians
that we may wish to compare our distribution to (which is which is not
really important.)  The distributions are plotted in terms of the 
number of standard deviations from the mean, where we have chosen the
standard deviation of the peak of the distribution as the most conservative
choice.  A fluctuation of $6-\sigma$ would correspond to PBH production
 under the Gaussian assumption, but the actual distribution under-produces
these fluctuations by $\sim 10^{150}$, resulting in almost no PBH production.}
\label{fig7}
\end{figure}

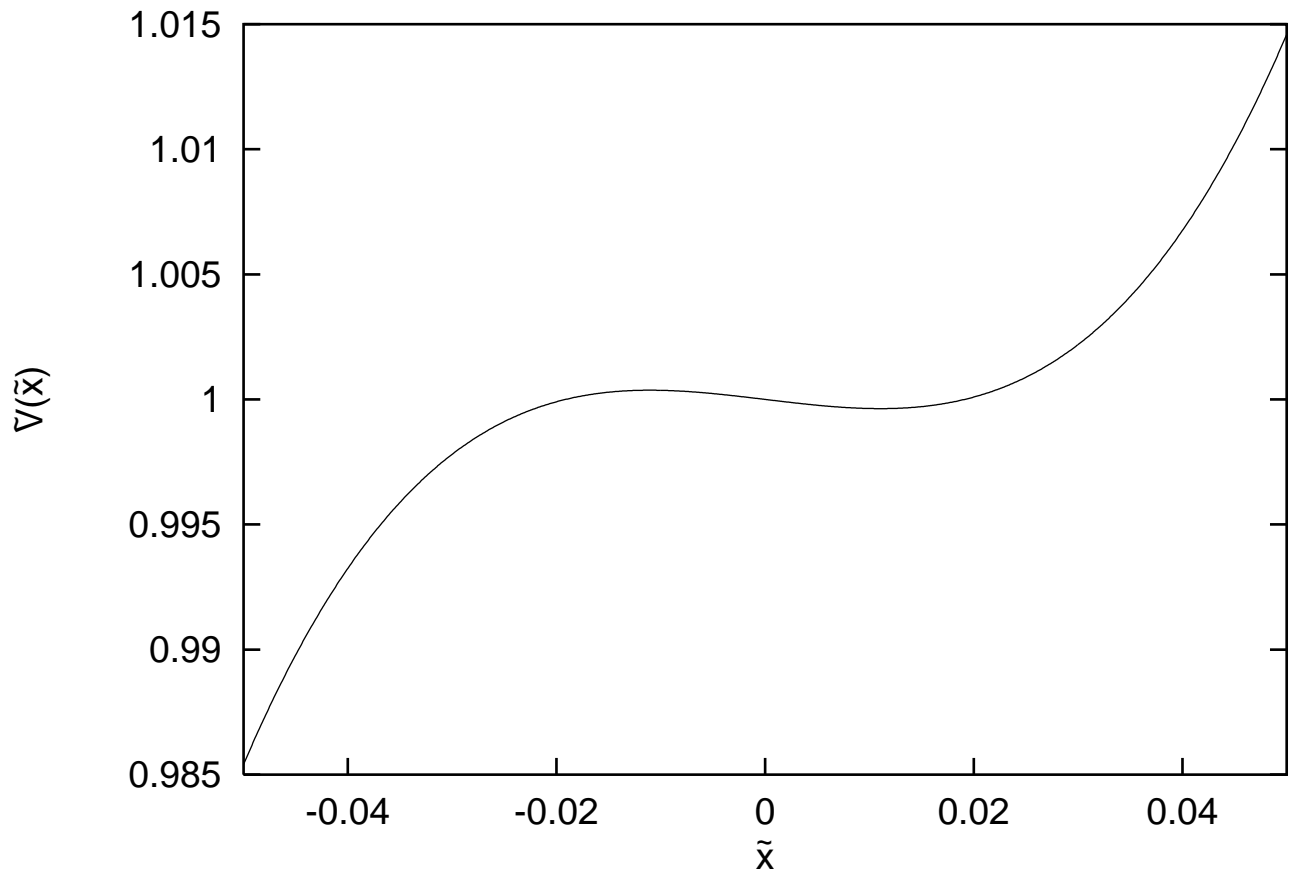
\begin{figure}
\caption{The wiggle potential (83): $\Vtw = 1 +
[136.717]x^{3} - 0.05x$.  The bump in the path
of the inflaton causes it to slow down, producing a spike in power
shown in Fig. 9.}
\label{fig8}
\end{figure}

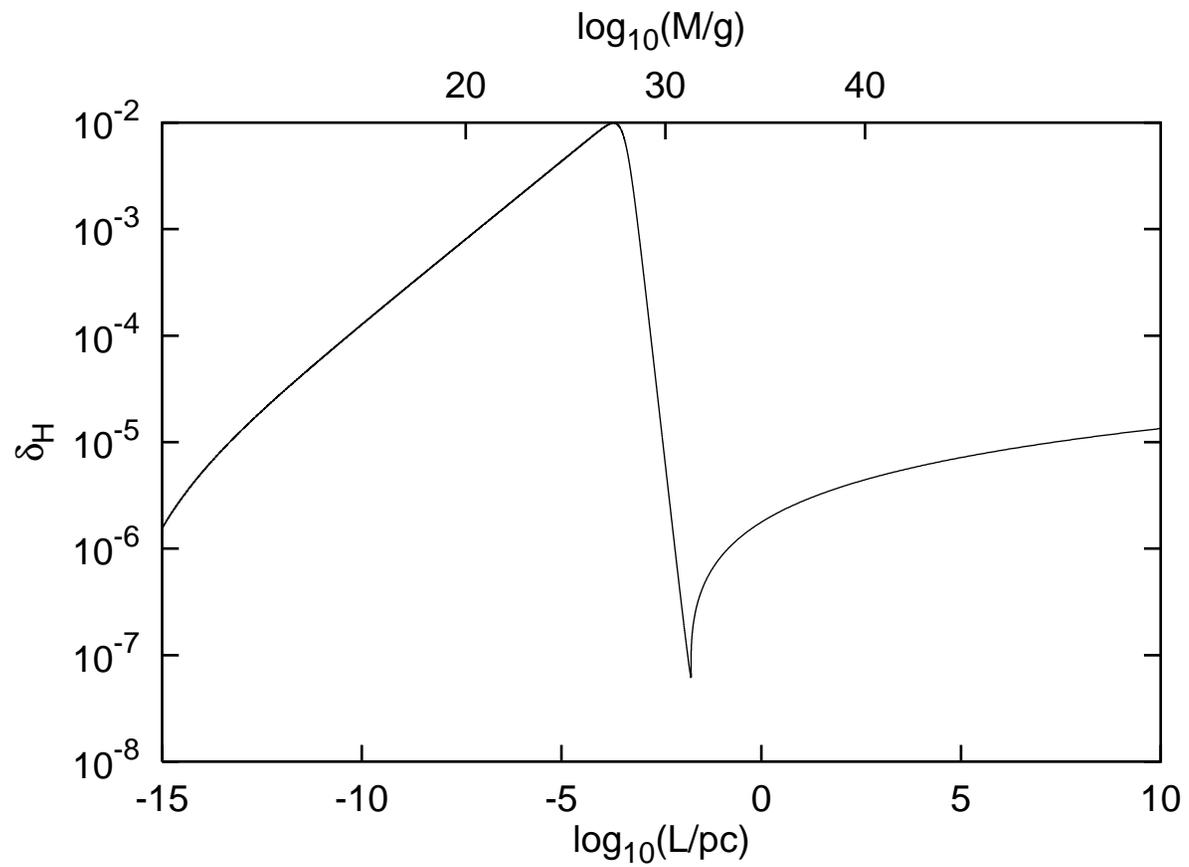
\begin{figure}
\caption{The spectrum of density fluctuations at horizon crossing,
 $\delH$, associated
with the wiggle potential (83).  The high-point in power corresponds 
to the production of $\sim 10^{28}$g PBH's.}

\label{fig9}
\end{figure}

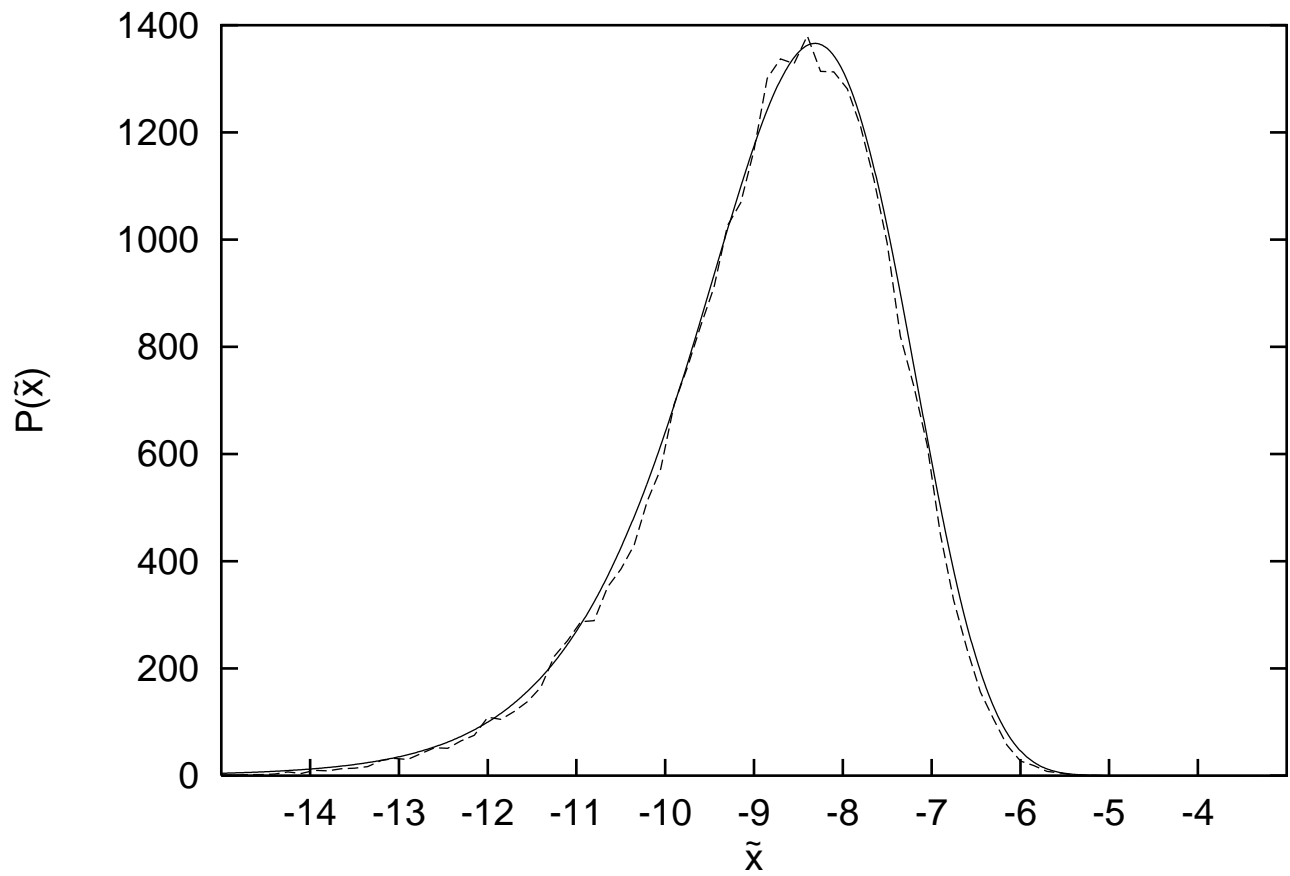
\begin{figure}
\caption{The solid line shows the calculated distribution
of fluctuations (94) associated with the wiggle model (83).  The dashed
line shows the results of our numerical simulation of this model, with
$4\times 10^{4}$ points.  The distribution (94) was normalized to the 
simulation number and bin size, and is clearly non-Gaussian
and skewed negative.}
\label{fig10}
\end{figure}

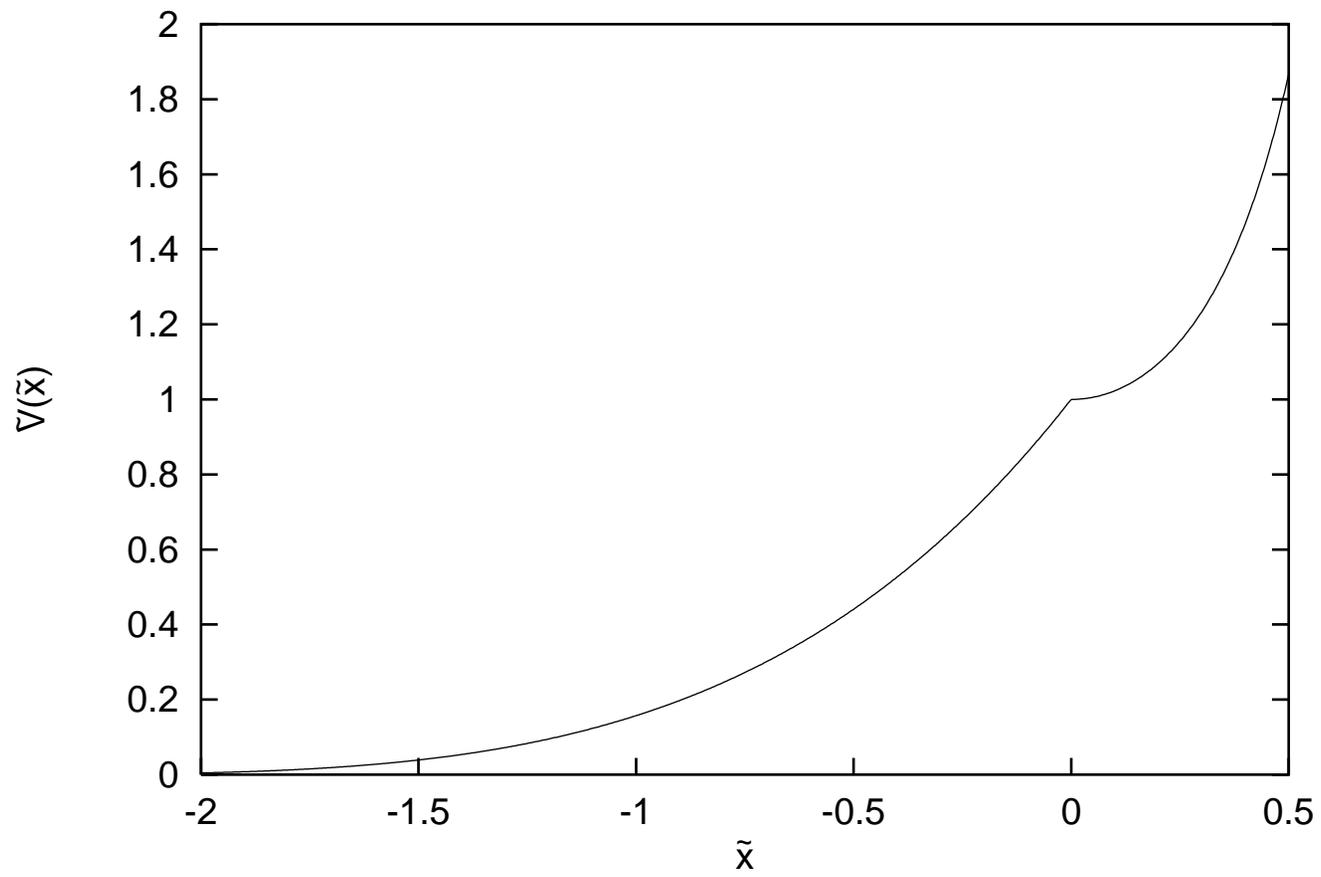
\begin{figure}
\caption{The cliff potential (95) which produces the blue spectrum 
shown in Fig. 12.  The final distribution
of fluctuations is nearly Gaussian since the slope of the potential has
a sharp increase just after the spike in power (see Fig. 12).}
\label{fig11}
\end{figure}

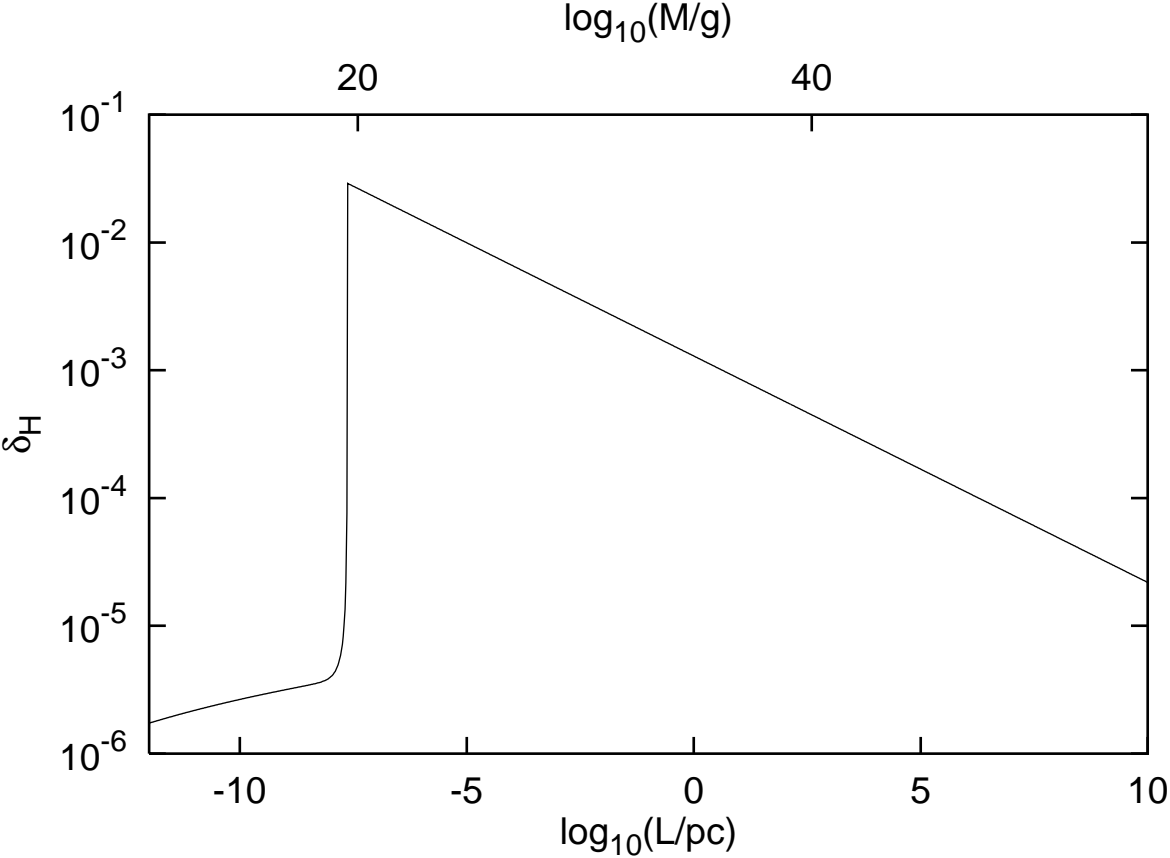
\begin{figure} 
\caption{The spectrum of density fluctuations at horizon crossing,
 $\delH$, associated with
the cliff potential (95).  The high-point in power corresponds to
the production of $\sim 10^{20}$g PBH's.}
\label{fig12}
\end{figure}


\end{document}